\documentclass[aps,prb,twocolumn,amsmath,amssymb,amsfonts,longbibliography,
    superscriptaddress,nobibnotes]{revtex4-2} 
\usepackage[utf8]{inputenc}
\usepackage[T1]{fontenc} 
\usepackage{ebgaramond}
\usepackage{braket}
\usepackage{graphicx} 
\usepackage{times}
\usepackage{soul}
\usepackage{url}
\usepackage{mathptmx}
\usepackage{gensymb}
\usepackage[
hypertexnames=false,
colorlinks=true,
citecolor=blue,
linkcolor=blue,
urlcolor=blue]
{hyperref}

\usepackage{float}

\usepackage{xcolor}
\definecolor{mygreen}{rgb}{0.0, 0.6, 0.0}
\DeclareMathAlphabet\mathbfcal{OMS}{cmsy}{b}{n}

\graphicspath{{fig/}{./fig/}{.}}

\begin{document}

\title{Interaction-driven transition between the Wigner crystal and the fractional Chern insulator in topological flat bands}

\author{Michał Kupczyński}
\email[e-mail: ]{michal.kupczynski@pwr.edu.pl}
\affiliation{Department of Theoretical Physics, 
Faculty of Fundamental Problems of Technology,
Wrocław University of Science and Technology,
PL-50370 Wrocław, Poland}

\author{Błażej Jaworowski}
\email[e-mail: ]{blazej@phys.au.dk}
\affiliation{Max-Planck-Institut f\"{u}r Physik komplexer Systeme, D-01187 Dresden, Germany}
\affiliation{Department of Physics and Astronomy, Aarhus University, DK-8000 Aarhus C, Denmark}

\author{Arkadiusz Wójs}
\email[e-mail: ]{arkadiusz.wojs@pwr.edu.pl}
\affiliation{Department of Theoretical Physics, 
Faculty of Fundamental Problems of Technology,
Wrocław University of Science and Technology,
PL-50370 Wrocław, Poland}

\date{\today}

\begin{abstract}
We investigate an interaction-driven transition between crystalline and liquid states of strongly correlated spinless fermions within topological flat bands at low density (with filling factors~$\nu=1/5$,~$1/7$,~$1/9$). Using exact diagonalization for finite size systems with periodic boundary conditions, we distinguish different phases, whose stability depends on the interaction range, controlled by the screening parameter of the Coulomb interaction. The crystalline phases are identified by a crystallization strength, calculated from the Fourier transforms of pair correlation density, while the Fractional Chern insulator phases are characterized using momentum counting rules, entanglement spectrum, and overlaps with corresponding Fractional Quantum Hall states. The type of the phase depends on a particular single particle model and its topological properties. We show that for~$\nu=1/7$ and~$1/5$ it is possible to tune between the Wigner crystal and Fractional Chern insulator phase in the kagome lattice model with the band carrying the Chern number~$C=1$. In contrast, in the $C=2$ models, the Wigner crystallization was absent at~$\nu=1/5$, and appeared at~$\nu=1/9$, suggesting that $C=2$~FCIs are more stable against the formation of crystalline order. 
\end{abstract}

\maketitle


\section{Introduction}
\label{sec.intro}

One of the most remarkable findings of the solid state physics in the last few decades, is the discovery of topological orders. 
They have changed the paradigm of matter phase classification and provided a potential way to construct a fault-tolerant quantum computer, based on the non-Abelian fractional (anyonic) statistics \cite{kitaev2003fault,sarma2005topologically}.

Among the most thoroughly studied examples of the topological orders are the fractional quantum Hall (FQH) states \cite{laughlin, jain1989composite, HaldaneHierarchy}. 
They were initially observed in the 2D electron gas in a strong magnetic field, where they appear at a fractional filling of the first or second Landau level \cite{tsui}. 
Alternative realizations were proposed in lattice systems. 
In this case, the role of a Landau level is played by a topological flat band, i.e. an energy band with a small dispersion and nonzero Chern number, and the FQH-like states arising in this setting are called fractional Chern insulators (FCIs) \cite{SunNature,Neupert,PRX,Zoology}. 
They were experimentally realized in moir\'e lattices in graphene in a strong magnetic field \cite{spanton2018observation}, and there are strong indications that they can also be created without an external magnetic field \cite{Ahmed.Luiu.2020}. 
Alternative realizations include the optical lattices \cite{cooper1999composite,sorensen2005fractional,palmer2006high,palmer2008optical,hafezi2007fractional,opticalflux1, dipolarspin1, dipolarspin2, nielsen2013local} or arrays of optical cavities \cite{maghrebi2015fractional,cho2008fractional,hayward2012fractional,anderson2016engineering,umucalilar2012fractional,umucalilar2012fractional,hafezi2013nonequilibrium,kapit2014induced}, which can be easier to control than electronic systems.

The existence of FCIs in simple lattice models 
of spinless fermions is now well established by many theoretical works \cite{SunNature,Neupert,PRX,Zoology,BeyondLaughlin,hierarchy, highchern1,highchern2,highchern3,highchern4,We,Andrews.Soluyanov.2020,Andrews.Mohan.2021}. 
The FCIs can exhibit several phenomena which are missing in the usual continuum FQH effect. 
The most striking is the possibility of obtaining bands with arbitrary Chern number~$C$. For~$|C|=1$ the FCIs are the lattice analogs of the well-known FQH states (Laughlin, Moore-Read etc.) \cite{SunNature,Neupert,PRX,Zoology,BeyondLaughlin,hierarchy}. 
However, in the case of~$|C|>1$, one finds a new series of states \cite{highchern1,highchern2,highchern3,highchern4,moller2015fractional,Andrews.Moller.2018}, which are a modified version of the multi-layer Halperin FQH states \cite{wu2013bloch}.

 In order to design experiments, it is important to determine the stable regions of the desired phase. For both FQH and FCI, one of the factors determining this stability is the competition with other phases, e.g. the charge order. Being essentially flat bands, the Landau levels allow for the existence of a Wigner crystal (WC). 
 In the presence of the long range Coulomb interaction, WC becomes lower in energy than the FQH states as the filling factor decreases \cite{YoshiokaFukuyamaHF,MakiZotos,LamGirvin,ZhuQMC,YiQMC,MaksymED,HutchinsonED, HaldaneED,ShibataYoshiokaDMRG}, which was confirmed experimentally \cite{Andrei, Kukushkin}.  
 The competition of FCIs with charge density waves was studied for large filling factors \cite{varney2010interaction,SingleParticleGrushin, KourtisTriangular, kourtis2017weyl,kourtis2014combined,LiFCIWC,kourtis2017symmetry,Wilhelm.Lang.2021}.
 The existence of such charge-ordered states depends on the commensuration with the lattice. 
 On the other hand, in our earlier work \cite{TFBWigner}, we have shown that analogously to FQH systems, Wigner crystals also emerge at a low filling factor of topological flat bands for certain types of long-range interaction. 
 In such cases, significant commensuration effects were absent. 
 In general, the existence of Wigner crystal depends on the filling factor and the type of interaction. 
 For example, in the Landau levels, one can use the Haldane pseudopotential formalism to construct a parent Hamiltonian for a Laughlin state at arbitrarily low filling \cite{HaldaneHierarchy}, for which a Coulomb interaction would generate a Wigner crystal. 
 Thus, by changing the pseudopotential parameters, one can trigger a transition from FQH state to WC at a constant filling factor \cite{HaldaneED} (some control of these parameters in an experiment can be exerted e.g. by changing the width of a quantum well \cite{thiebaut2015fractional}).
 Inspired by these findings, we investigate the stability of various liquid and crystal phases, and transitions between them, in different lattice models.

In this work, we study fractionally filled topological flat bands in the presence of a density-density interaction with a screened Coulomb (Yukawa) potential by utilizing the exact diagonalization (ED) method to compute the spectra and eigenstates for finite-size systems, with periodic boundary conditions. Our main results are following: (i)~it is possible to trigger a transition between FCI and WC on bands with $C=1$ at fillings $\nu=1/7$ and $\nu=1/5$ by varying the range of the interaction, (ii)~the FCIs at $C=2$ are more stable against WC formation than the FCIs at $C=1$, (iii)~nevertheless it may be possible to observe a WC-FCI transiton in $C=2$~bands at filling~$\nu=1/9$.

The article is organized as follows. In Section~\ref{sec:methods} we describe the~$C=1$ and~$C=2$ tight-binding models used by us, as well as the  details of the exact diagonalization procedure. Next, in Section~\ref{sec:1over7}, we show that for~$\nu=1/7$ of a~$C=1$ band, the long- and short-range interaction give rise to respectively WC and FCI phases, we introduce the WC and FCI characteristics and study them as a function of interaction range. Then, in Section~\ref{sec:C2}, we compare the~$C=1$ and~$C=2$ systems at filling~$\nu=1/5$, showing that the former display a WC-FCI transition, while in the latter we observe the FCI for all considered interaction ranges. Also, we study the~$C=2$ bands at~$\nu=1/9$, and observe both the FCI and WC phases, although their behaviour as a function of the interaction range strongly depends on the system size and chosen lattice model. The section~\ref{sec:conclusions} summarizes the results.


\section{Models and methods}\label{sec:methods}

\subsection{Lattice models}\label{ssec:models}

 We consider various tight-binding models which exhibit a non-zero Chern number of the lowest band. Within an energy band, the crystal momentum eigenstates are given by the Bloch wavefunctions,
 \begin{equation}
    \psi_\mathbf{k}(\mathbf{r})=e^{i\mathbf{k}\cdot \mathbf{r}} u_{\mathbf{k}}(\mathbf{r}),
 \end{equation}
 where $u_{\mathbf{r}}(\mathbf{r})$ is lattice-periodic. The Chern number is defined as the integral of Berry curvature divided by $2\pi$, i.e.
 \begin{equation}
     C=\frac{i}{2\pi} \iint_{BZ} \left[ \braket{\frac{\partial u_{\mathbf{k}}}{\partial k_y}|\frac{\partial u_{\mathbf{k}}}{\partial k_x}}- \braket{\frac{\partial u_{\mathbf{k}}}{\partial k_x}|\frac{\partial u_{\mathbf{k}}}{\partial k_y}}
     \right]\mathrm{d}k_x\mathrm{d}k_y
     \label{eq:chern}
 \end{equation}
where BZ denotes the Brillouin zone. 
 The Chern number \eqref{eq:chern} is proportional to the Hall conductivity of the fully filled band.
 
\begin{figure}
    \centering
    \includegraphics[width=0.5\textwidth]{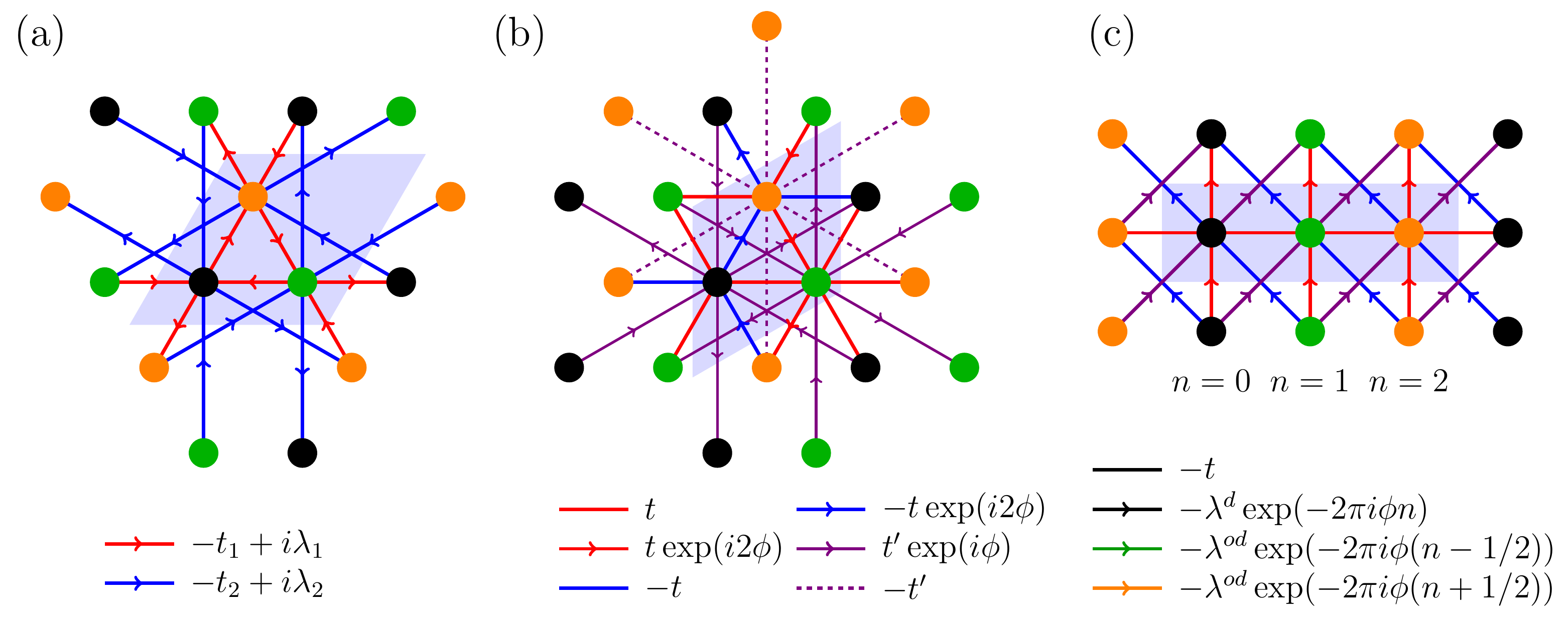}
    \caption{The lattice models used in our work: (a)~kagome lattice (b)~triangular lattice  (c)~generalized Hofstadter model on square lattice. Only the hoppings starting and/or ending within a given unit cell (purple parallelograms) are shown.}
    \label{fig:lattices}
\end{figure}

 As an example of a model with~$C=1$ of the lowest band, we take the kagome lattice described by the Hamiltonian \cite{Tang,Zoology}
\begin{equation}
H_{\mathrm{kag}}=-\sum _{\langle i,j\rangle }({t}_{1}\pm {\rm{i}}{\lambda }_{1}){c}_{i}^{\dagger }{c}_{j}-\sum _{\langle \langle i,j\rangle \rangle }({t}_{2}\pm {\rm{i}}{\lambda }_{2}){c}_{i}^{\dagger }{c}_{j}, 
\label{eq:hkag}
\end{equation}
where $c_{i} ({c}_{i}^{\dagger})$ anihilates (creates) a particle on $i$-th lattice site, $\langle \rangle $, $\langle \langle \rangle \rangle $ denote nearest and next-nearest neighbours, respectively. The ``$+$'' corresponds to the hoppings along the arrows in Fig.~\ref{fig:lattices}~(a) and the ``$-$'' to the hoppings in the opposite direction. At $\lambda_1=\lambda_2=t_2=0$, the model is gapless, with one band being exactly flat. 
Introducing nonzero $\lambda_1$ creates a pattern of effective magnetic flux, which is zero on the average, but breaks the time-reveresal symmetry, which is necessary for nonzero Chern numbers. At $\lambda_2=t_2=0$, the middle band has $C=0$, and the two other bands have opposite Chern numbers with $|C|=1$ (except from $\lambda_1=0$ and $\lambda_1=\pm \sqrt{3}t_1$ where the model is gapless). Inclusion of nonzero second-neighbour hoppings $t_2$, $\lambda_2$ allows to tune the band dispersion. Although the lowest band has~$C=1$ in a wide range of parameters, it is nearly flat (resembling a Landau level) only in a certain part of this range. Moreover, even if we disregard the single-particle energies in the ED computation (see Sec. \ref{ssec:ed}), making the bands artificially flat, not all parameter values are favourable for FCIs, due to e.g. the fluctuations of the Berry curvature \cite{Zoology}. Considering systems with $\nu=1/5$~filling, we use the parameters $t_1=1$,~$t_2=-0.3$,~$\lambda_1= 0.28$ and $\lambda_2=0.2$, corresponding to a nearly flat lowest band with $C=1$ \cite{Tang}, which shown to host a fermionic FCI phase at $\nu=1/3$ \cite{Zoology}. As we will show later, the $\nu=1/5$ FCI can also exist there. In the case of~$\nu=1/7$, for which the FCI phase is much less stable than for higher fillings, we keep $t_1=1$ and $t_2=-0.3$, but we use $\lambda_1= 0.5$ and~$\lambda_2=0.2$, for which allow to increase the FCI stability (see Appendix~\ref{app:params}).

In addition to the kagome lattice, we also study two models with a $C=2$ lowest band. In general, the $|C|>1$ models  \cite{highchern3, trescher2012flat} can be created systematically by stacking several layers of $|C|=1$ models. However, the first model we use, the triangular lattice model, was found independently from this method \cite{wang2012fractional}. It is defined by the Hamiltonian
\begin{equation}
H_{\mathrm{tri}}=\pm t\sum_{\langle i,j\rangle}\exp(i\phi_{ij}) c^\dagger_i c_j  \pm t'\sum_{\langle\langle i,j\rangle \rangle} c^\dagger_i c_j \exp(i\phi_{ij}').
\label{eq:htri}
\end{equation}
Here, ``$+$'' and ``$-$'' refer to the hopping denoted by solid and dashed lines in Fig.~\ref{fig:lattices}~(b), respectively. Each of the phases has three possible values, $\phi'_{ij}\in \{-2\phi, 0, 2\phi\}$, $\phi_{ij}\in \{-\phi, 0, \phi\}$, where the positive sign refers to the hopping along the arrows in Fig.~\ref{fig:lattices}~(b), negative sign to the hopping in the opposite direction, and 0 to the hoppings without an arrow. Following Ref.~\cite{wang2012fractional}, we choose $t=1$,~$t'=1/4$,~$\phi=\pi/3$, for which we obtain a nearly-flat lowest band with $C=2$, while the two other bands have $C=-1$ each. At these values of parameters, the lowest band can host bosonic FCIs at different fillings \cite{wang2012fractional,highchern4}, thus we expect that the same will happen for fermions.

The second $C=2$ model is a generalized Hofstadter model on the square lattice, i.e. a Hofstadter model with second-neighbor hoppings \cite{wang2013tunable},
\begin{multline}
H_{\mathrm{Hof}}=-\sum_{n,m}\big( t c^\dagger_{n, m}c_{n+1, m}+\lambda^{\mathrm{od}}e^{2\pi i \phi (n+\frac{1}{2})}c^\dagger_{n, m}c_{n+1, m+1}+\\+
\lambda^{\mathrm{od}}e^{-2\pi i \phi (n+\frac{1}{2})}c^\dagger_{n, m}c_{n+1, m-1}+
\lambda^{\mathrm{d}}e^{2\pi i \phi n}c^\dagger_{n, m}c_{n, m-1}
\big).
\label{eq:hhof}
\end{multline}
Here, instead of labeling sites with a single index~$i$, we label them with two indices~$n,m$ denoting their~$x$ and~$y$ positions in the lattice, respectively. The model is highly tunable. At $\lambda^{\mathrm{od}}=0$, $\lambda^{\mathrm{od}}=t$
it reduces to an ordinary Hofstadter model on square lattice \cite{Hofstadter}. At rational values of the flux, $\phi=a/b$, with $a\in\mathbb{Z}$, $b\in\mathbb{N}^{+}$ and $a,b$ coprime, the model has $b$ bands. At small flux, these bands are a lattice approximation of the continuum Landau levels, and thus have $|C|=1$. At higher flux, the Landau level structure is no longer visible, but the bands still are topologically nontrivial. The second-neighbour hopping $\lambda^{\mathrm{od}}$ can mix them, leading to topological phase transitions and changes in the Chern numbers. Ref. \cite{wang2013tunable} shows that even in the relatively simple case of $\phi=1/3$ one can obtain a rich phase diagram, with Chern number up to $|C|=4$ in the middle band and up to $|C|=2$ in two other bands. In this work, we use $\phi=1/3$, which results in the unit cell containing three sites. The other parameters are fixed at $t=1$,~$\lambda^{\mathrm{d}}=1$,~$\lambda^{\mathrm{od}}=-1/2$, which leads to the presence of a nearly-flat $C=2$ lowest band, which was shown to host a $\nu=1/3$ bosonic FCI \cite{jaworowski2019characterization}.

\subsection{Interaction Hamiltonian}\label{ssec:ed}
We study finite size systems with  $N_1\times N_2$
unit cells along two real space lattice vectors and $N_{\mathrm{part}}$~particles. We impose periodic boundary conditions, so the total momentum is a good quantum number. We consider two-body interaction in a form of the screened Coulomb (Yukawa) potential
\begin{equation}
    \hat{V}=\sum_{i,j}V(r_{ij})n_in_j
    \label{eq:interaction}
\end{equation}
where $n_i$, $n_j$ are the particle densities at~$i$ and~$j$ sites, $r_{ij}$~is the smallest distance between the sites~$i$ and~$j$ on the torus (i.e. with periodic boundary conditions taken into account). 
The interaction potential is
\begin{equation}
    V(r)=\exp\left(-\alpha\left(r-r_{NN}\right)\right)/r,
  \label{eq:yukawa}
\end{equation}
where $r_{NN}$ is the distance between the nearest-neighboring sites, and $\alpha$ is the screening parameter that will be varied to trigger the phase transition. 
We note that in general, instead of considering only the closest periodic image of a given site, we could have calculated the sum of contributions of all its periodic images, however, since we consider strong screening, we expect that the differences between these two approaches will be small. Without the loss of generality, we set~$r_{NN}=1$.

To focus only on the interaction effects and to reduce the computational complexity of our calculations, instead of the full Hamiltonian, we diagonalize \begin{equation}
    H_{\mathrm{int}}=P\hat{V}P,
    \label{eq:projint}
\end{equation}
where $P$ is the operator of projection to the lowest band. Thus, we first diagonalize the single-particle Hamiltonians given by Eqs.~\eqref{eq:hkag},~\eqref{eq:htri},~\eqref{eq:hhof}, and then construct the many-particle basis by distributing $N_{\mathrm{part}}$~particles over the momentum eigenstates in the lowest band. In order to focus on interaction effects, we use the flat band approximation by neglecting the band dispersion. For a given filling factor~$\nu= \frac{N_{\mathrm{part}}}{N_1 N_2}$, the resulting Hamiltonian matrix is diagonalized using the Implicitly Restarted Arnoldi Method implemented in the ARPACK package.

\section{ Phase transition between crystal and liquid phases at filling~$\nu=1/7$ }\label{sec:1over7}

To study the crystal and liquid phases, we need indicators which have large values when the given phase is stable and small (or vanishing) values when the phase is absent. 
In subsections~\ref{ssec:WCsignatures} and~\ref{ssec:FCIsignatures}, these indicators are introduced on the example of~$N_1\times N_2 = 5\times 7$ plaquette of the kagome lattice at filling factor~$\nu=1/7$.
Finally, in subsection~\ref{ssec:1over7transition} the phase transition between the liquids and crystal phases at filling~$\nu=1/7$ is analysed.

\subsection{The crystal phase for C=1 band at~$\nu=1/7$}
\label{ssec:WCsignatures}

We begin with the long-range limit of screened Coulomb interaction, in which the Wigner crystallization is expected. The crystalline properties of many-body eigenstates~$|\psi\rangle$ are determined using the pair correlation density (PCD), defined as
\begin{equation}
G(i,j)=\frac{\braket{\psi|c^{\dagger}_{i}c^{\dagger}_{j}c_{j}c_{i}|\psi}}{\braket{\psi |c^{\dagger}_{i}c_{i}|\psi}}.
\end{equation}
For future analysis, we replace every site with a Gaussian function and make the PCD continuous (see Appendix~ \ref{app:cryst_strength} for details).

Fig.~\ref{fig:wc}(a) shows the many-body spectrum for~$\alpha=0.5$. 
The energy differences are very small and decrease as the screening increases (which means that distant particles interact less strongly),
thus the energies are renormalized through division by the next-nearest neighbour interaction strength~$V_{NNN}$ (in kagome lattice, $V_{NNN}=V(\sqrt{3})$). 

Fig. \ref{fig:wc}(b) shows the PCD  of the ground state. 
The PCD in the plaquette is shown together with its periodic repetitions. 
The peaks of the PCD and the fixed particle indicated by red triangles form an almost-hexagonal lattice. 
Within the plaquette, there are four peaks, which correspond to the four particles, plus one fixed particle giving~$N_{\mathrm{part}}=5$.

The periodicity of the Wigner crystal can be characterized by the peaks in the Fourier transform of PCD either in Cartesian or polar coordinates, similarly as it has been done in our previous work \cite{TFBWigner}. The Cartesian transform is performed along two real space lattice vectors, yielding the quantity~$F_{mn}$, where $m, n$ are integers describing the momenta (see Appendix~\ref{app:cryst_strength} for definition).
For the comparison of different plaquettes, it is convenient to normalize~$F_{mn}$ by the~${F}_{00}=N_{\mathrm{part}}-1$. We define~$\tilde{F}_{mn}$ as~$\tilde{F}_{mn} = F_{mn}/F_{00}$. To avoid the effects of the periodicity related to the periodic repetition of the considered plaquette, only the Fourier peaks~$\tilde{F}_{mn}$ of~$m$ and~$n$ smaller than plaquettes sizes are taken into account.
In Fig.~\ref{fig:wc}(c) we plot the magnitude of the normalized Cartesian Fourier coefficients~$|\tilde{F}_{mn}|$.
A clear reciprocal lattice, with a unit cell smaller than in the reciprocal lattice of the underlying kagome lattice, is seen as brighter peaks around the peak at zero. This is a necessary condition for the presence of the Wigner crystals. The magnitude of these Fourier peaks decay as we move away from~$m=0$,~$n=0$, which is a consequence of particles having a finite spatial extent (see Ref.~\cite{TFBWigner} for details).

We note that the Fourier transform of PCD is in fact much less anisotropic than it looks in Fig. \ref{fig:wc}(c) at the first glance. The apparent anisotropy comes from the fact that the Fourier transform is performed in the direction of the two primitive vectors of the kagome lattice, which are not orthogonal –- the angle between them is 60\degree. The $m$ and $n$ integers describe the Fourier components in these directions – in other words, they describe the coordinates of the points in reciprocal space along the reciprocal lattice vectors, the angle between which is 120\degree. Thus, if we plotted Fig. \ref{fig:wc}(c) in true reciprocal space, with 120{\degree} angle between the axes, and with $n$ and $m$ rescaled according to Eq. \ref{eq:pcd_reciprocal}, the Fourier peaks would be arranged in a lattice much closer to hexagonal (see Fig. 1 in \cite{TFBWigner}). However, plotting $\tilde{F}_{mn}$ as in \ref{fig:wc}(c) makes it easier to determine $m$ and $n$ of the peaks.

At a given particle number~$N_{\mathrm{part}}$, there is only a finite number~$N_W$ of possible Wigner lattices, each characterized by two Fourier components at momenta~$(m_i, n_i)$ and~$(o_i, p_i)$, $i=1,2,\dots N_W$, corresponding to the two fundamental vectors of the reciprocal Wigner lattice. 
To obtain periodicity in both directions, both of these components should be nonzero. 
We define the crystallization strength as the square product of the magnitudes of these two components, normalized by the zeroth momentum component, maximized over all possible crystals
\begin{equation}
    W=\max_{i\in [1,N_W]} \sqrt{|\tilde{F}_{m_i n_i}||\tilde{F}_{o_i p_i}|}.
    \label{eq:cryst_strength}
\end{equation}
In contrast to our previous work, Ref.~\cite{TFBWigner}, the square root is added to the definition of~$W$ to obtain a magnitude of~$W$ comparable to a single Fourier peak~$\tilde{F}_{mn}$. Ideally,~$W$ should be~0 for a perfectly flat PCD and~1 for an array of Dirac deltas (i.e. point-like particles). However, because of the existence of ``exchange-correlation hole'' around the fixed particle, the PCD is never perfectly flat even for liquids, and thus even liquids can have small nonzero~$W$. 
Nevertheless, the transition between a liquid and a crystal should be accompanied with by an increase of~$W$. 
The crystallization strength~$W$ is shown as a color scale in Fig.~\ref{fig:wc}(a) and one can notice that a set of the lowest energy states are of a crystalline character.

Alternatively, we can look at polar coordinates and obtain the transform~$\tilde{F}_{\theta}(r,k_\theta)$ in the angular direction only (see Appendix~\ref{app:cryst_strength} for details). Here, $r$~is the distance to the fixed particle and $k_{\theta}$~is an integer describing momentum in the angular direction. To define angular crystallization strength, let us first define the peak strength at given~$k_{\phi}$ as the normalized magnitude of the Fourier component maximized over all possible values of the radius
\begin{equation}
    \tilde{F}_{\mathrm{peak}}(k_\phi)=\max_{r<r_{\mathrm{max}}}|\tilde{F}_{\theta}(r,k_\theta)|,
\end{equation}
where $r_{\mathrm{max}}$ is defined in Appendix~\ref{app:cryst_strength}. The angular crystallization strength is defined as ~$W_\theta=\max_{k_\theta=2,4,6} \tilde{F}_{\mathrm{peak}}(k_\theta)$. The~$W_\phi$ alone is not sufficient to determine the existence of the crystal, as $2-$fold symmetry is exhibited also e.g. by the stripe order. On the other hand, it probes not only the existence of WC but also its symmetry. 

The angular Fourier transform is shown in Fig.~\ref{fig:wc}(d). The range~$r$ corresponds to the red circle indicated in Fig.~\ref{fig:wc}(b). There is a maximum of angular density at~$k_\theta=0$ (the zeroth Fourier component) around maximal~$r$, which  corresponds  to  the  six  peaks closest to the fixed particle. At this radius, we observe also a relatively strong $k_\theta=6$~Fourier component, showing that the PCD is approximately six-fold rotationally-symmetric, i.e. close to the hexagonal lattice. Note that we also have a nonzero Fourier component at $k_\theta=4$ (and, weaker, at $k_\theta=2$), which occurs because the Wigner crystal is not perfectly hexagonal, as the perfectly hexagonal Wigner lattice is not permitted by the boundary conditions for a $5\times 7$ system.

\begin{figure}
    \centering
    \includegraphics[width=0.5\textwidth]{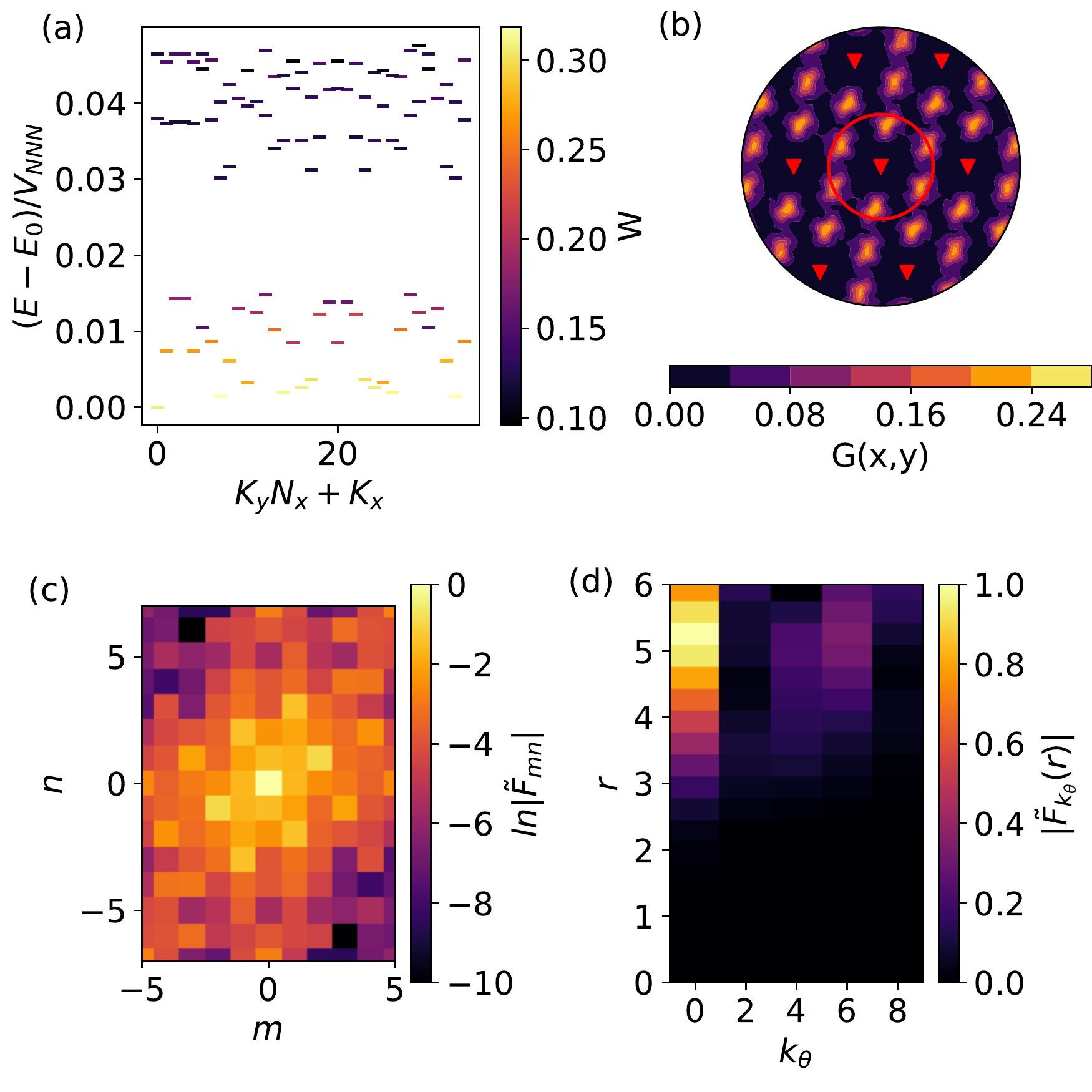}
    \caption{
    Wigner crystal in a $N_1\times N_2 = 5\times 7$ kagome system with $N_{\mathrm{part}}=5$ particles for the screening parameter~$\alpha = 0.5$. (a)~Energy spectrum normalized by the interaction between next-nearest-neighbor interaction~$V_{NNN}$. The color scale indicates Cartesian crystallization strength. (b)~Pair correlation density of the ground state (located in~$\mathbf{K}=[0,0]$ subspace). The plaquette is drawn together with periodic images to make the crystal structure more visible. The red triangles denote a fixed particle in the center and its periodic images. (c)~The Cartersian Fourier transform of the PCD from~(b). (d)~The polar Fourier transform of the PCD from~(b). Only the even components are drawn. The upper limit of the plot corresponds to the red circle indicated in~(b). 
    }
    \label{fig:wc}
\end{figure}

\subsection{The liquid phase for C=1 band at $\nu=1/7$}\label{ssec:FCIsignatures}

In the limit of short-range interaction the FCI state is expected as the ground state. To study this state, we fix the parameter as~$\alpha = 6.0$.

\begin{figure}
    \centering
    \includegraphics[width=0.5\textwidth]{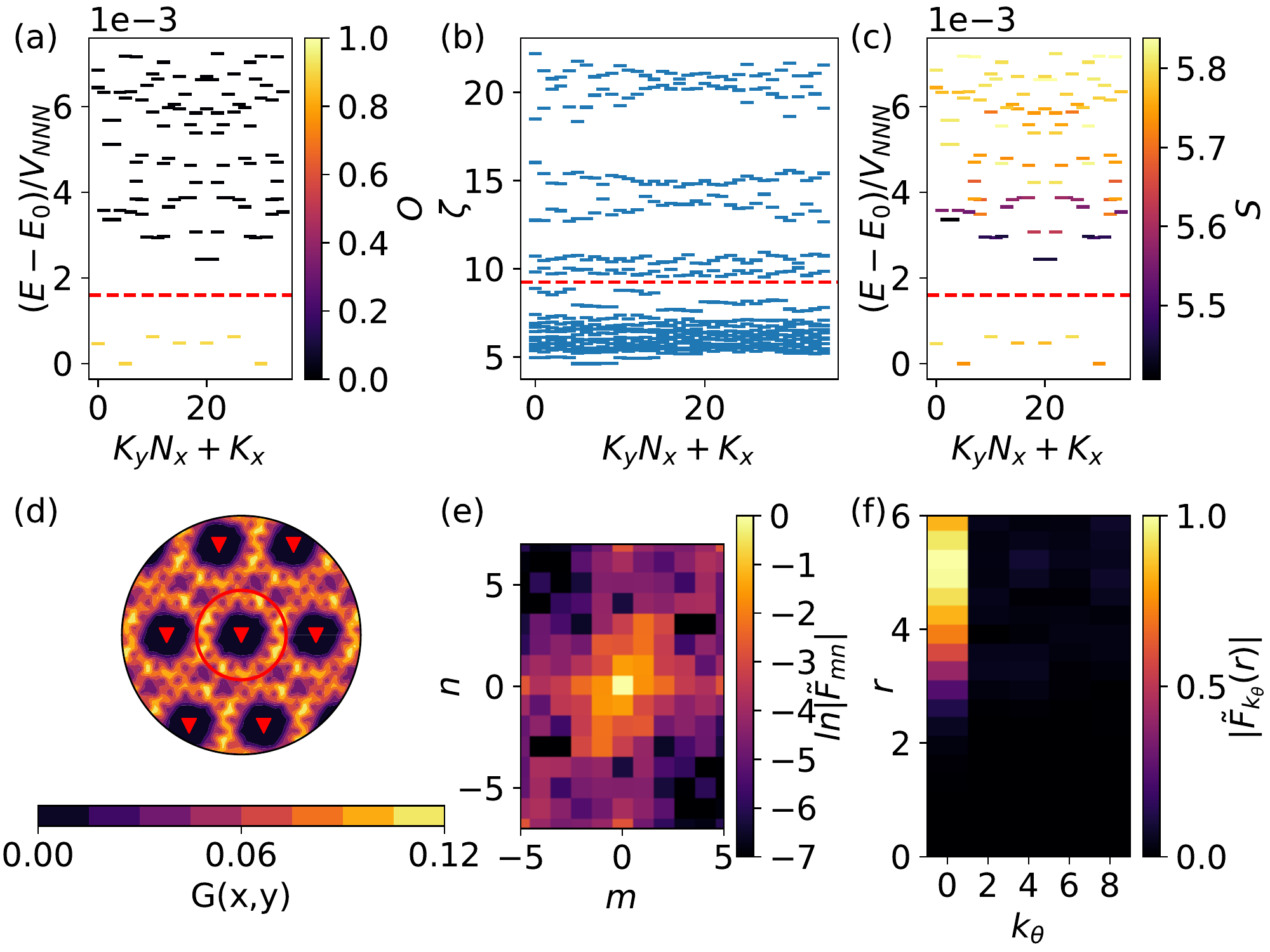}
    \caption{
    FCI in a $N_1\times N_2 = 5\times 7$ kagome system with $N_{\mathrm{part}}=5$~particles for the screening parameter~$\alpha = 6.0$. (a)~Energy spectrum normalized by the interaction between next-nearest-neighbor interaction~$V_{NNN}$. The color scale indicates overlap~$O$ with model FQH states. The overlaps in the  momentum  subspaces that do not correspond to a model FQHE ground state are set to~0 by a definition. (b)~Entanglement spectrum of the absolute ground state (located in the~$\mathbf{K}=[0,6]$ subspace) obtained after tracing out all but $N_A=2$~particles. (c)~Energy spectrum normalized by the next-nearest-neighbor interaction~$V_{NNN}$. The color scale indicates entanglement entropy~$S$. 
    (d)~Pair correlation density of the absolute ground state. The plaquette is drawn along with periodic images to make the crystal structure more visible. The red triangles denote the periodic images of the fixed particle. (e)~The Cartersian Fourier transform of the PCD from~(d). (f)~The polar Fourier transform of the PCD from~(d). Only the even components are drawn. The upper limit of the plot corresponds to the red circle shown in~(d). 
    }
    \label{fig:fci}
\end{figure}

The fractional Chern insulator phase is identified by looking at various signatures of topologically nontrivial liquid state \cite{PRX,bernevig2012emergent,li2008entanglement,sterdyniak2011extracting,haque2007entanglement,Zozulya2,wu2013bloch}. 
For the Laughlin states at filling~$\nu=1/q$,~$q\in \mathbb{N}^+$, we expect $q$~quasi-degenerate states separated by a gap from the rest of the spectrum. 
The momenta of these states are determined by the appropriate generalized Pauli principle \cite{bernevig2012emergent, PRX}. 
Fig.~\ref{fig:fci}(a) shows the seven nearly degenerated states separated by the energy gap to excited states.
The momenta of the quasi-degenerate ground states agree with predictions from the generalized Pauli principle of $\nu=1/7$~Laughlin FCI \cite{PRX,bernevig2012emergent}.
To avoid confusion, we note that throughout this work, in the cases where we observe quasi-degeneracy, we will use the phrase ``quasi-degenerate ground states'' to refer to the entire manifold and ``absolute ground state'' to refer to the lowest-energy one.

The ground state momentum counting rule is not a definite proof of FCI existence. It should be supplemented e.g. by the analysis of the particle entanglement spectrum \cite{li2008entanglement,sterdyniak2011extracting}, which should reveal a nonzero gap $\Delta \zeta$ between the low energy sector with an appropriate number of states below the gap in each momentum sector that agree with the appropriate generalized Pauli principle. 
While typically in the literature one constructs the density matrix as an equally-weighted superposition of pure state density matrices of all the quasi-degenerate ground states, we construct it from a single ground state (see Appendix~\ref{app:entropy}), which turns out to be sufficient to obtain the correct FCI entanglement energy level counting. The entanglement spectrum is obtained after tracing out all but $N_A=2$ particles (we use~$N_A=2$ for all the systems investigated in this work).
In Fig.~\ref{fig:fci}(b) $\Delta \zeta$~is denoted by the red dash line with a correct number of states below the gap confirming FCI. We note that there are also more gaps higher in the entanglement spectrum. They are absent for model FQH states, where the lowest gap is infinite. For FCIs, some of these gaps were connected with another generalized Pauli principles, which may reflect different types of correlations that can be generated by the Hamiltonian \cite{Zoology}. However, for the identification of the type of topological order in the given state, only the lowest gap is relevant.

We also calculate the entanglement entropy~$S$, computed in the particle partition (the same as for the entanglement spectrum). 
The numerical value for~$S$ can be compared with exact bounds, $S_{\mathrm{max}}$~--~the largest entropy permitted by the generalized Pauli principle, and~$S_{\mathrm{min}}$~--~the entropy of a single Slater determinant. For FQHE, $S$ was shown to be close to the former limit \cite{haque2007entanglement,Zozulya2}, and we expect that the same will happen for FCI. 
Although, as we will see later, this approach is less reliable than the entanglement spectrum, in some cases it does detect the transition between FCI and WC. The definitions of entanglement-related quantities can be found in Appendix~\ref{app:entropy}.
In Fig.~\ref{fig:fci}(c), entanglement entropy is shown as the color scale on the energy spectrum. The quasi-degenerated ground state is characterized by the high entropy values. 
For the investigated system, the lower and upper bounds for the entanglement entropy are~$S_{\mathrm{min}}\approx 2.30$ and~$S_{\mathrm{max}}\approx 5.95$ (see Appendix~\ref{app:entropy} for the definitions). The entropy for all states in the ground state manifold is close to the upper limit, as in the FQH systems \cite{haque2007entanglement,Zozulya2}.

We also compare the overlaps~$O=|\braket{\psi|\psi_{FQH}}|^2$ between the state~$\ket{\psi}$ and the ground state~$\ket{\psi_{FQH}}$ of a model FQH Hamiltonian within the same momentum subspace. The gauge is fixed according to the prescription from Ref.~\cite{wu2013bloch}. The color scale in Fig.~\ref{fig:fci}(a) shows corresponding overlaps which are $O > 0.88$ for the seven quasi-degenerate ground states. 
The comparison is performed only for the momentum subspaces, in which there is a model ground state. For all the others, we simply assign a zero overlap. 
More details of these calculation can be found in Appendix~\ref{app:overlap}.

 Fig.~\ref{fig:fci}(d) shows the PCD of the absolute ground state at a limit of short range interactions for $\alpha=6$. A nearly uniform PCD, apart from the vicinity of the fixed particle, indicates it is a liquid state. 
 This is confirmed by the Cartesian Fourier spectrum shown in Fig.~\ref{fig:fci}(e) and the polar Fourier spectrum shown in Fig.~\ref{fig:fci}(f), which do not show any clear reciprocal Wigner lattice, thus the state is approximately rotationally and translationally invariant.

\subsection{Phase transition between crystal and liquid phases for C=1~band at~$\nu=1/7$ }\label{ssec:1over7transition}

In the two previous subsections, we have shown a few signatures which allow for distinguishing liquid and crystal phases. 
In the example system, the FCI state is a true ground state in the limit of the short-range interaction, and many low-energy states are Wigner crystals in the limit of the long-range interaction.
Thus, the phase transition between liquid and crystal is expected by tuning $\alpha$~parameter in the screened Coulomb interaction \eqref{eq:interaction}. 

\begin{figure}
    \centering
    \includegraphics[width=0.5\textwidth]{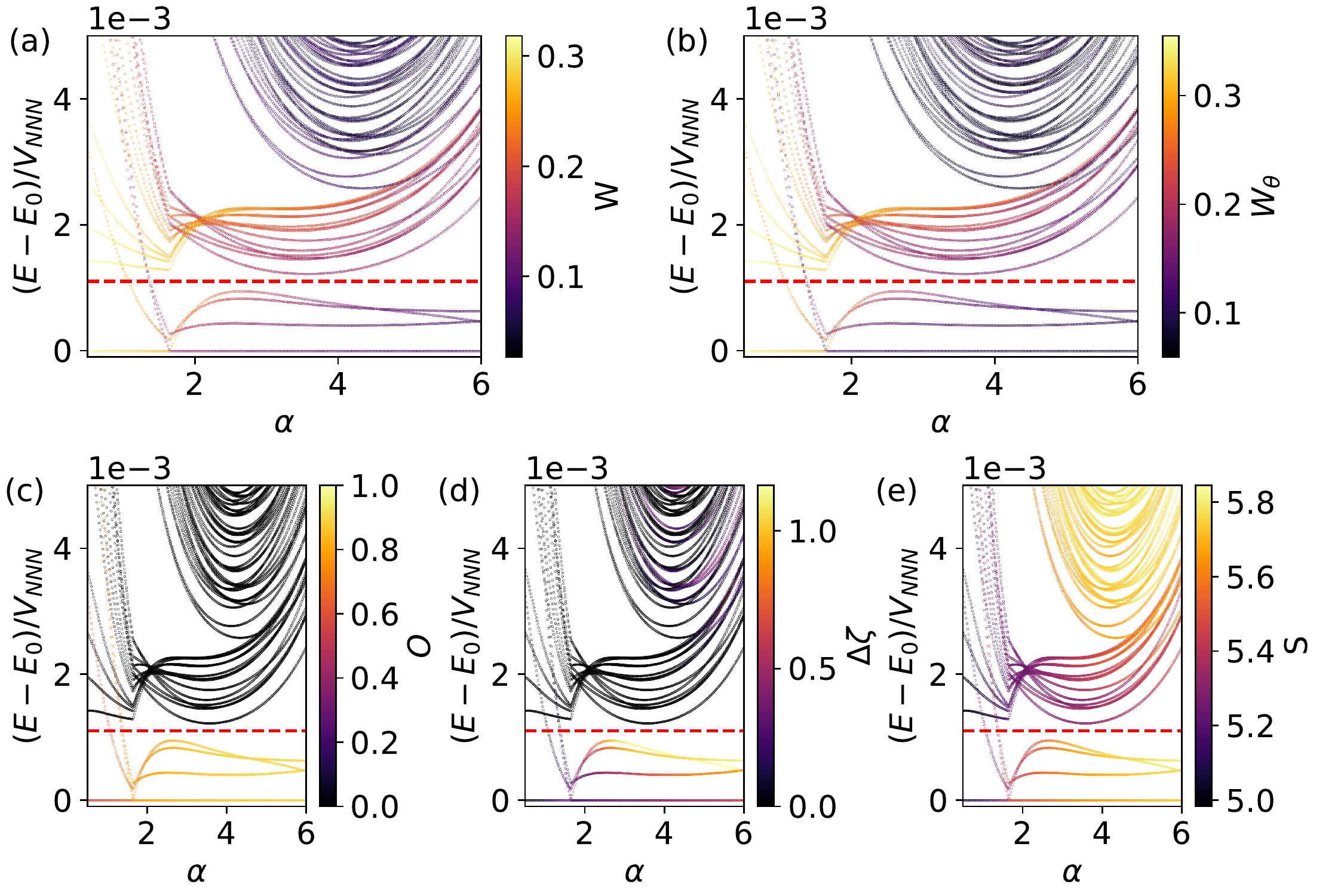}
    \caption{
    The energy spectrum of a~$5\times 7$ kagome system as a function of $\alpha$, normalized by the next-nearest-neighbor interaction~$V_{NNN}$. The color denotes Wigner crystallization indicators in the upper row: (a)~Cartesian crystallization strength, (b)~polar crystallization strength, and FCI indicators in the lower row: (c)~overlap with model FQH states, (d)~the gap in the entanglement spectrum, (e)~entanglement entropy.
    In~(c), the overlaps in the momentum subspaces not corresponding to a model FQHE ground state are set to 0 by definition.
    }
    \label{fig:WC_FCI_17}
\end{figure}

We begin from studies of the phase transition on the previously considered kagome plaquette $N_1\times N_2 = 5\times 7$ at the filling~$\nu=1/7$. 
Fig.~\ref{fig:WC_FCI_17} shows the evolution of the energy spectrum as a function of~$\alpha$, measured with respect to the absolute ground state energy, with crystallization and liquid signatures indicated by color scales. 

The color scale in Fig.~\ref{fig:WC_FCI_17}(a) shows the Cartesian crystallization strength $W$ and the polar crystallization strength $W_\theta$ in Fig.~\ref{fig:WC_FCI_17}(b).
For low $\alpha$, there is a single ground state with relatively large~$W$ and~$W_{\theta}$. 
Low-lying excited states also display a crystalline order with even larger~$W$ and~$W_{\theta}$, compared to the ground state crystallization strength. 
As the interaction range is decreased (larger~$\alpha$), crystallization strength~$W$ and~$W_{\theta}$ decreases for all states. 

Above~$\alpha \approx 1.32$ the seven states with the lowest energy become separated from the rest of the spectrum by the gap indicated by a red dashed line.
The momenta of these states agree with predictions from the generalized Pauli principle of $\nu=1/7$~Laughlin FCI \cite{PRX}.
The energy split between that states is minimized at~$\alpha \approx 1.65$, and leads to the level crossing at this point. 
These seven states are characterized by different crystallization strengths~$W$ and~$W_{\theta}$.
After the crossing, for larger~$\alpha$, the crystallization strength of low-energy states is small, but there are some states above the energy gap, which display relatively large~$W$ and~$W_{\theta}$ even above~$\alpha=2$. However, for sufficiently large $\alpha$ the crystalline order disappears from all the states.

The quasi-degenerate states at large $\alpha$ are indeed the FCI states, which was confirmed by calculating the FCI signatures in Fig~\ref{fig:fci}(c)-(e).
The overlap between the energy eigenstates and the model FQHE ground states is denoted in Fig.~\ref{fig:WC_FCI_17}(c). 
The seven states forming the ground state manifold have overlap with the model states $O > 0.55$ in the entire range of~$\alpha$, even below the gap closure at~$\alpha \simeq 1.32$, but reach~$O > 0.88$ in the limit of large~$\alpha$. 
The overlap of the model ground state with the excited states is close to zero. 

The evolution of the gap in the entanglement spectrum~$\Delta\zeta$  is shown as a color scale in Fig.~\ref{fig:WC_FCI_17}(d). It can be seen that this gap is open for the seven quasi-degenerate ground states at large~$\alpha$, and decreases (eventually vanishing), as $\alpha$~decreases. In contrast, in the excited states the entanglement gap is much smaller or nonexistent.

The  evolution of entanglement entropy of the states is shown as a color scale in Fig.~\ref{fig:WC_FCI_17}(e). 
For the investigated system, the lower and upper bounds for the entanglement entropy are~$S_{\mathrm{min}} \approx 2.30$ an~$S_{\mathrm{max}}\approx 5.95$ (see Appendix~\ref{app:entropy} for the definitions). 
For~$\alpha \approx 6$ the entropy for all states is~$S\approx 5.8$, which is close to the upper limit, as in the FQH systems \cite{haque2007entanglement,Zozulya2}. 
As $\alpha$ is lowered and the system undergoes the transition to WC, the entropy decreases. The entropy can therefore be a good signature of the FCI. Nevertheless, even for crystalline states it remains well above the minimal value corresponding to a single Slater determinant. While some excited states have similar values of entropy as the ground states we note that the bound $S_{\mathrm{max}}$ is valid only for the ground states, so the comparison with this bound does not tell us anything about the nature of excited states. We also notice that the low-lying excited states (some of which exhibit crystalline order even when the FCI ground state manifold is fully formed) have significantly lower entanglement entropy than the quasi-degenerate ground states. Also, we observe that in the entire energy spectrum the entanglement entropy decreases with decreasing $\alpha$.

\begin{figure}
    \centering
    \includegraphics[width=0.5\textwidth]{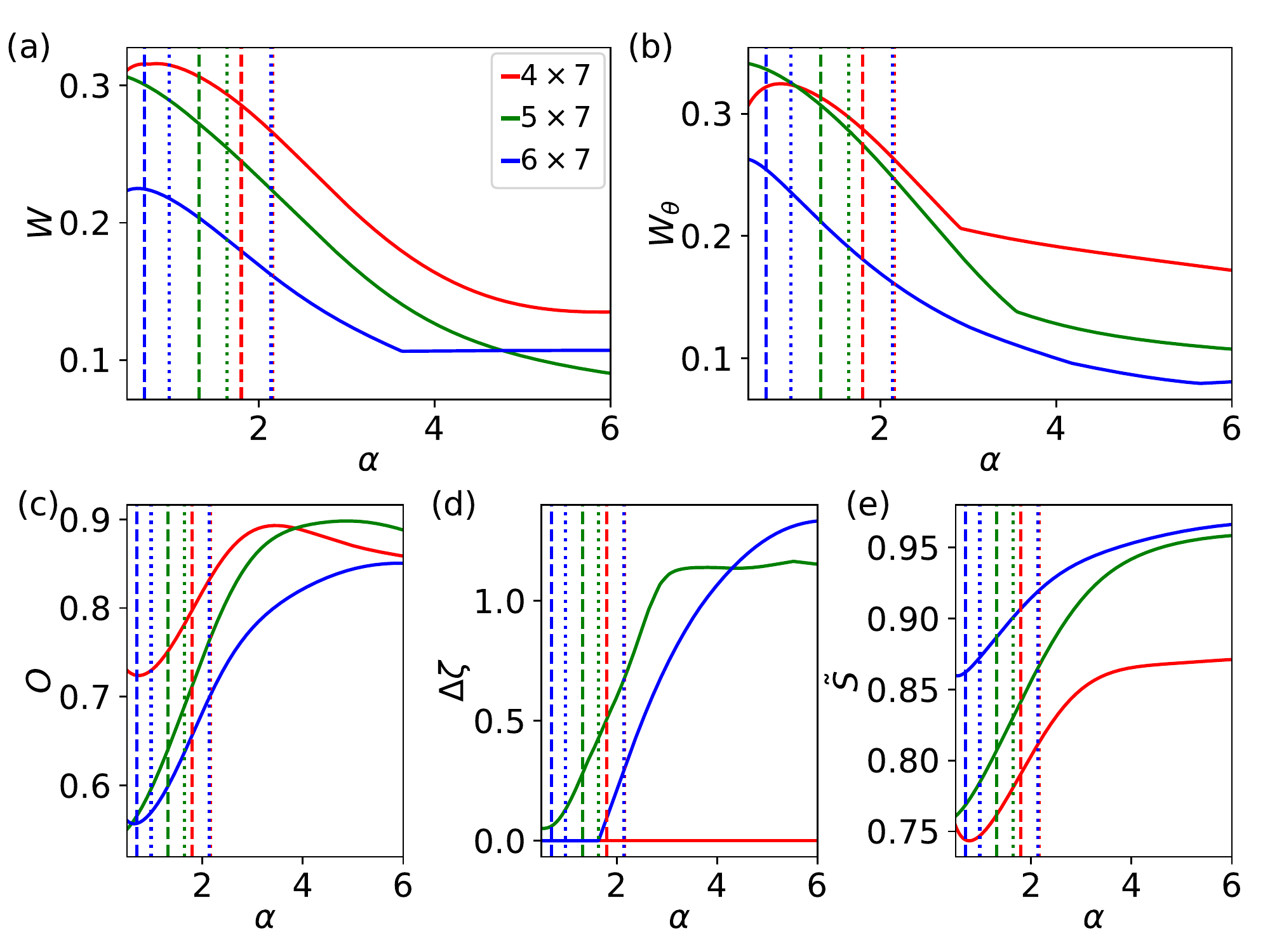}
    \caption{The FCI and WC characteristics as a function of~$\alpha$ for three kagome systems with~$C=1$ at filling~$\nu=1/7$ (a)~Cartesian crystallization strength, (b)~polar crystallization strength, (c)~overlap with model states, (d)~entanglement gap, (e)~renormalized entanglement entropy~$\tilde{S}=(S-S_{\mathrm{min}})/(S_{\mathrm{max}}-S_{\mathrm{min}})$.
    The dashed vertical lines denote the $\alpha$~values for which the gap above the seven quasi-degenerate FCI states closes. The dotted vertical lines correspond to the location of the characteristic crossing of all seven quasi-degenerate states (there are two such crossings for the~$6\times 7$ system, see Fig. \ref{fig:different_sizes_1_7}).}
    \label{fig:WC_FCI_17_all}
\end{figure}

We can conclude that the crystalline phase is stable for the long-range interactions, and is replaced by the FCI phase in the limit of the short-range interaction on the considered $5\times 7$ kagome plaquette.
For the analysis of finite size effects, the signatures of both phases are plotted in Fig.~\ref{fig:WC_FCI_17_all} for three kagome plaquettes with sizes $4\times 7$,~$5\times 7$,~$6\times 7$ at filling~$\nu=1/7$. We do not consider a~$7\times 7$ system, as on $N_1=N_2$ plaquettes the degeneracy of the crystal may prevent its detection \cite{TFBWigner}.
Because these characteristics behave differently for each state, here we plot them for a state selected in the following way: we choose the momentum subspace in which the absolute ground state is located at low~$\alpha$ (e.g.~$\mathbf{K}=[0,0]$ for the $5\times 7$ system), and then we select the lowest state from this subspace at each $\alpha$.
Because this subspace fulfills the FCI generalized Pauli principle, the state in question turns into one of the states from quasi-degenerate FCI manifold as $\alpha$~increases.
The full spectrum with the WC and FCI indicators is shown in the appendix~\ref{app:1over7}.
For every considered system, the crystallization strength, both Cartesian $W$ (Fig.~\ref{fig:WC_FCI_17_all}(a)) and polar $W_{\theta}$ (Fig.~\ref{fig:WC_FCI_17_all}(b)), decreases when $\alpha$ increases from values around $0.3-0.2$ in the crystal phase to, below $0.2$ in the liquid phase. 
In the limit of the short-range interaction for all plaquettes seven quasi-degenerated states are separated by the energy gap from the rest of the spectrum.
Momenta of that ground states are in full agreement with the counting rules for the Laughlin state~$\nu=1/7$.
Moreover, the increase of overlap~$O$ with FQHE state (Fig.~\ref{fig:WC_FCI_17_all}(c)), and normalized entanglement entropy (Fig.~\ref{fig:WC_FCI_17_all}(e)) when crystallization strengths decrease, proves that the the crystal phase is replaced by the FCI phase.
The similar behavior is visible in the gap in the entanglement spectrum (Fig.~\ref{fig:WC_FCI_17_all}(d)) for plaquettes $5\times 7$ and~$6\times 7$.
The gap in the entanglement spectrum is not visible for the chosen state in the~$4\times 7$ system, but it exists in a few other quasi-degenerate ground states, so that the value of the average gap over all FCI states is non-zero.
It indicates that the Fractional Chern Insulator in that system is not as stable in the smallest systems, as in the bigger ones.
It is important to notice here, that calculating entanglement spectrum for only one state instead of the superposition of all FCI states is not the standard approach, so the lack of the gap in the entanglement spectrum is not equivalent to the lack of the FCI state.

Previous subsections identify the crystal and the liquid phases in the limit of small~$\alpha$ (long range interaction, a crystal limit) and large~$\alpha$ (short range interaction, a liquid limit). In Figs. \ref{fig:WC_FCI_17} and \ref{fig:WC_FCI_17_all}, one can notice that that the phase transition does not occur abruptly, it is rather continuous: the phase indicators change smoothly and can remain relatively large even when a given phase fully vanishes. However, this lack of sharp jumps may be a result of the small size of investigated systems. 
Moreover, the fact overlap remain relatively high in the WC phase  strongly suggest that FCI states have some crystal-like correlations built in. 
Similar behaviour was reported for analogous phase transition in FQHE models \cite{HaldaneED}.

Because the changes in most FCI/WC characteristics are gradual, it is hard to define a transition point. Definitions using some threshold on the values of these indicators would always be arbitrary. Another way to define the transition point is to look at characteristics which can take only the binary “yes” or “no” values, for example the existence of the energy gap above the seven states described by generalized Pauli principle. The level crossing leading to the closure of this gap occurs in all three systems, at $\alpha \approx 1.8$, $\alpha \approx 1.32$, $\alpha \approx 0.7$ for $4\times 7$, $5 \times 7$ and $6 \times 7$ systems, respectively (they are denoted by dashed vertical lines in Fig. \ref{fig:WC_FCI_17_all}). While from these results there seems to be a general trend of the transition point moving to the lower $\alpha$ with increasing system sizes, in such small systems, strong finite size effects prevent us from estimating the transition point in the thermodynamic limit.

We note that while our results are limited to the kagome lattice model and a certain set of parameter values, in our previous paper we have shown that the stability of the Wigner crystal does not strongly depend on the model \cite{TFBWigner}. Thus, the observed phase transition should be visible in any model for any set of parameter values, for which the FCI phase exists for short-range interaction. Moreover, these phenomena should not be limited to the considered screened Coulomb interaction, but should be visible in other similar types of density-density interaction, which allows manipulation of the interaction range. 

\section{Crystal-liquid phase transition on the $C=2$ models} \label{sec:C2}

In the previous section, the phase transition between Wigner crystal and Fractional Chern Insulator phases has been analysed at filling~$\nu=1/7$ on the flat band with Chern number $C=1$.
In this section, the results are extended to the models with Chern number~$C=2$ at~$\nu=1/5$ and~$\nu=1/9$ (in the former case, we also compare it with a~$C=1$ band at the same filling factor). 
The~$C>1$ FCIs can be understood as multilayer FQH states with ``color-entangling'' boundary conditions, which mix the layers \cite{highchern1,highchern2, wu2013bloch,wu2014haldane}. 
For example, the~$\nu=1/5$ and~$\nu=1/9$ FCIs on~$C=2$ bands, studied in this section, are modified Halperin states at filling $\nu=2/5$ and~$\nu=2/9$, respectively (the difference in fillings is a consequence of different definitions of filling factor for FCI and FQHE).

The indicators of crystal and liquid phases at~$C=2$ are the same as for~$C=1$. Counting rules are more complicated comparing to the~$C=1$ case \cite{wu2014haldane}, but instead of implementing them, we simply compare the momenta of the ground states (or entanglement energy levels) with results for a model FQH system \cite{wu2013bloch,wu2014haldane}. These model states can also be used to compute overlaps.

\subsection{Topological phase transition at $\nu=1/5$ for C=1 and C=2 bands}
\label{ssec:1_5}

\begin{figure}
    \centering
    \includegraphics[width=0.5\textwidth]{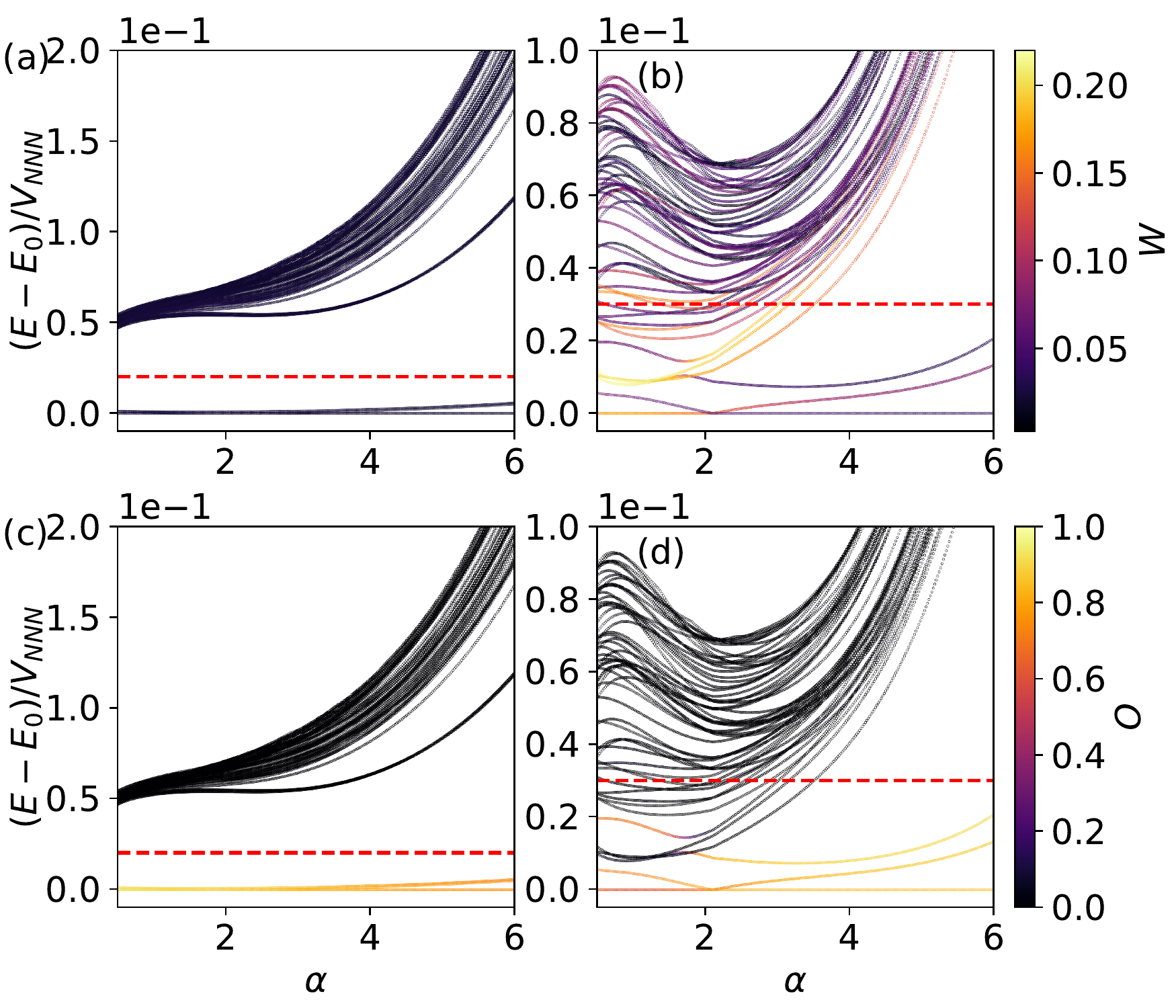}
    \caption{ The energy spectrum of~$6\times 5$ $C=1$ and~$C=2$ plaquettes as a function of~$\alpha$, normalized by the next-nearest-neighbour interaction~$V_{NNN}$ at filling $\nu = 1/5$. 
    The left column corresponds to the triangular lattice system with the flat band with Chern number~$C=2$ (plots~(a) and~(c)), the right column corresponds to the kagome system with the flat band with Chern number~$C=1$  (plots~(b) and~(d)).
    The color denotes Cartesian Wigner crystallization strength in the upper row (plots~(a) and~(b) ),
    and overlap with the model FQHE state in the lower row (plots~(c) and~(d) ). 
 The overlaps in the momentum subspaces not corresponding to a model FQHE ground state are set to 0 by definition.
 }
    \label{fig:WC_FCI_15_compare}
\end{figure}

The FCIs in~$C=2$ bands do not exist at~$\nu=1/7$. Therefore, to compare the~$C=1$ and~$C=2$ cases, we study the~$\nu=1/5$ filling. In the case of~$C=2$ at~$\nu=1/5$, 
Fig.~\ref{fig:WC_FCI_15_compare}(a) shows a low-energy spectrum as a function of the $\alpha$ parameter on the~$6 \times 5$ triangular lattice plaquette with $N_{\mathrm{part}}=6$ particles. The energy gap between five low energy states and higher energy states is visible in the whole range of $\alpha$ parameter. 
Momenta of these states agree with the momenta of the FQH states, which strongly suggests that the system is in the FCI phase. The color scale denotes strength of the Cartesian Fourier transform of the Wigner crystallization~$W$.
The low values of~$W < 0.05$ mean that none of the quasi-degenerate ground states is a Wigner crystal. The lack of crystallization is confirmed by visual inspection of the PCDs.
To prove that the system is in the FCI phase, the overlap of the energy eigenstates with the FQHE ground states is shown in Fig.~\ref{fig:WC_FCI_15_compare}(c). It is high for the ground state on the whole $\alpha$ range. Its value is the highest in the limit of the long-range interactions, and its minimum value is equal to approximately~$O \approx 0.75$ in the limit of short-range interaction. 
In conclusion, in the considered system the WC phase does not exist, and the FCI state is more stable for the long-range interaction than in the short-range limit.

The results for Chern number~$C=2$ at filling~$\nu = 1/5 $ are compared with the system with the same filling but with Chern number~$C=1$. 
In Fig.~\ref{fig:WC_FCI_15_compare}~(a) the crystallization strength $W$ is presented on the low-energy spectrum of~$6\times 5$ kagome plaquette with $N_{\mathrm{part}} = 6$ particles, and Fig.~\ref{fig:WC_FCI_15_compare}~(c) shows the overlap with model FQHE states for the same system.
The Cartesian crystallization strength is about~$W \approx 0.17$ in the absolute ground state at low~$\alpha$ (which is doubly degenerate in this case) and even higher in some excited states, indicating the presence of crystalline order. The five quasi-degenerate ground states become separated from the rest of the spectrum at~$\alpha \approx 1.80$. The momenta of these states fully agree with the counting rules. The ground states become FCI in the limit of the short-range interaction, confirmed by calculating the overlap with the FQHE states, which achieves~$O \approx 0.9$.
Thus, the phase transition between FCI and WC phases, exist at~$C=1$ at filling~$\nu=1/5$, and is similar as the one occurring at~$\nu=1/7$. This indicates that the lack of phase transition in the system with Chern number~$C=2$ is not an effect of the filling~$\nu=1/5$ only, and it suggests that it could be an effect of the Chern number value.

\begin{figure}
    \centering
    \includegraphics[width=0.5\textwidth]{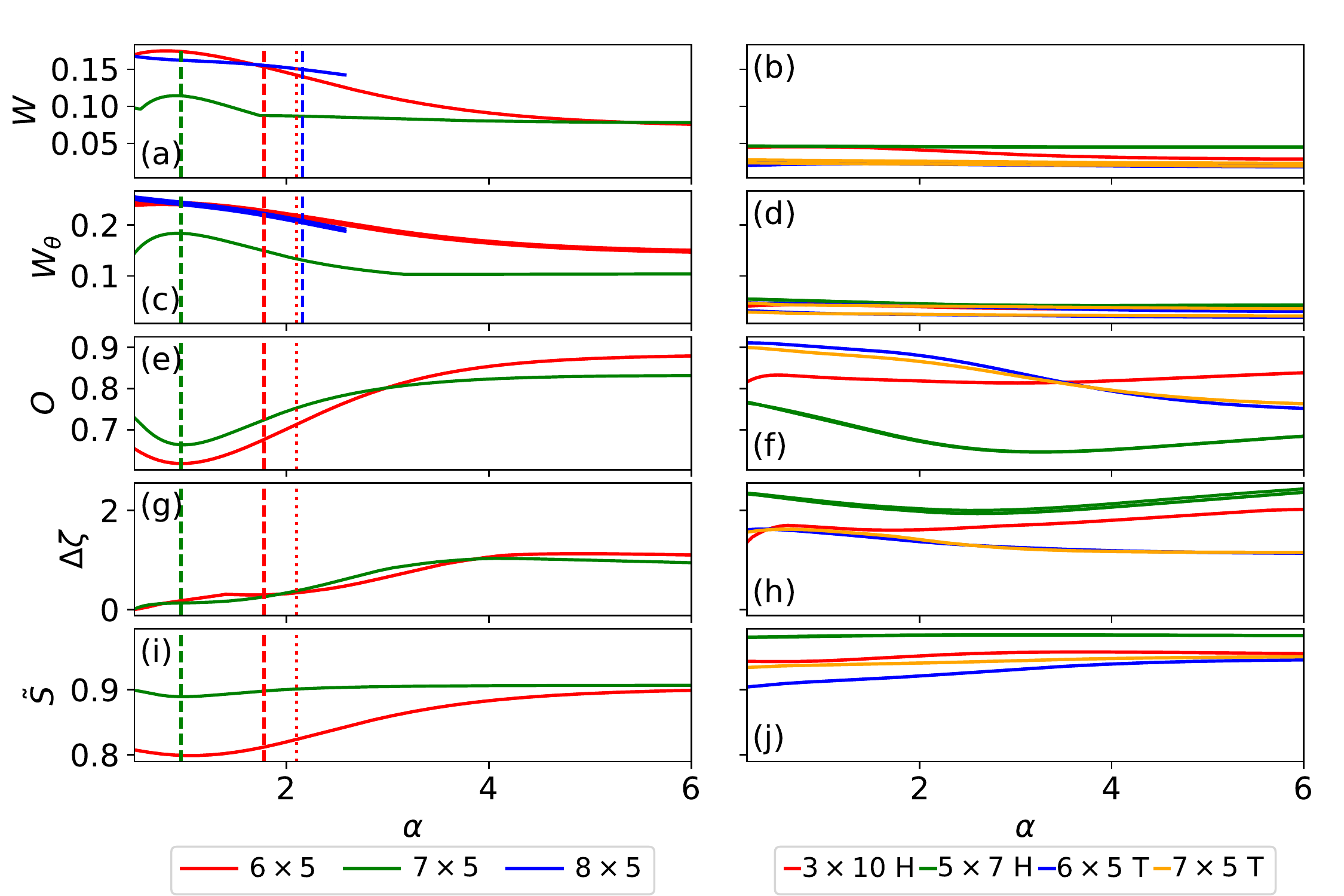}
    \caption{The FCI and WC characteristics as a function of $\alpha$ for different models and plaquettes at filling $\nu=1/5$ for the one chosen state.
     The left column corresponds to the kagome lattice with the flat band with Chern number~$C=1$ (plots: (a), (c), (e), (g), (i)), the right column corresponds to the triangular lattice model (T) and Hofsdtater model (H) with the flat band with Chern number~$C=2$ (plots: (b), (d), (f), (h), (j)).
     The different plaquettes sizes are denoted by different colors.
     In the first row the Cartesian crystalization strentgh $W$ is shown (plots~(a) and~(b)), in the next row shows polar crystalization strength $W_{\theta}$ (plots~(c) and~(d)).
     In the last three rows the FCI characteristics are presented: the overlap with model FQHE state (plots~(e) and~(f)), the entanglement gap (plots~(g) and~(h)), and the renormalized entanglement entropy~$\tilde{S}=(S-S_{\mathrm{min}})/(S_{\mathrm{max}}-S_{\mathrm{min}})$ (plots~(i) and~(j) ).
    The dashed vertical lines denote the~$\alpha$ values for which the gap above the five quasi-degenerate FCI states closes. The dotted vertical line corresponds to the location of the characteristic crossing of all five quasi-degenerate states.
     The results for the~$8\times 5$ plaquette are obtained only for some characteristics and only in a limited range of~$\alpha$, because of the numerical complexity of the considered system and problems with numerical convergence of diagonalization problem, especially in the limit of high $\alpha$ values. }
    \label{fig:WC_FCI_15_combine}
\end{figure}

We study the phases at filling~$\nu = 1/5$ for different plaquettes and models (including the generalized Hofstadter model). 
The results for a single, selected state are shown in Fig.~\ref{fig:WC_FCI_15_combine}. 
The state is chosen in the same way as in Fig.~\ref{fig:WC_FCI_17_all}. If there is exact degeneracy, the degenerate states display similar characteristics, so just one subspace is chosen. 
The full spectrum for these other systems with the WC and FCI indicators is shown in the appendix~\ref{app:1over5} for the case with $C=1$ and in the appendix~\ref{app:1over5_2} in the case with $C=2$.
From Fig.~\ref{fig:WC_FCI_15_combine} we can see that the phase transition occurs in systems with the Chern number~$C=1$ and does not occur in considered systems with~ $C=2$. The crystallization strength for~$C=1$ is, in general, smaller for~$\nu=1/5$ than for~$\nu=1/7$. This shows that the Wigner crystals at~$\nu=1/5$ are more fragile than for~$\nu=1/7$, which is in line with the results from \cite{TFBWigner}, showing that the crystallization strength increases as the filling is lowered (although we note that here we use different single-particle parameters at $\nu=1/5$ and~$\nu=1/7$, see Sec. \ref{ssec:models}). The gap closing for $C=1$ occurs at $\alpha\approx 1.78$, $\alpha\approx 0.96$,  $\alpha\approx 2.16$ for $6\times 5$, $7\times 5$ and $8\times 5$ systems, respectively (see the dashed vertical lines in Fig. \ref{fig:WC_FCI_15_combine}). Contrary to the $\nu=1/7$ case, here we do not observe any clear trend in the location of the transition point as a function of the system size, which may be due to geometric effects.

The results presented in this subsection show, that the considered phase transition and the stability of the crystal strongly depend on the value of the Chern number.

\subsection{Topological phase transition at~$\nu=1/9$ for~$C=2$ band}
The next available filling for a fermionic $C=2$~FCI is~$\nu=1/9$.
For such a case, we study the following systems: $4\times 9$ and~$6\times 6$, both with~$N_{\mathrm{part}}=4$, for the generalized Hofstadter model, and $4\times 9$,~$6\times 6$ and~$5\times 9$ with $N_{\mathrm{part}}=4$,~$N_{\mathrm{part}}=4$ and~$N_{\mathrm{part}}=5$, for the triangular lattice model.
The~$N_{\mathrm{part}}=5$ case is not considered for the Hofstadter model, as for rectangular Bravais lattice the WCs are degenerate on the classical level, which can prevent their detection using~$W$ or~$W_\theta$.

\begin{figure}
    \centering
    \includegraphics[width=0.5\textwidth]{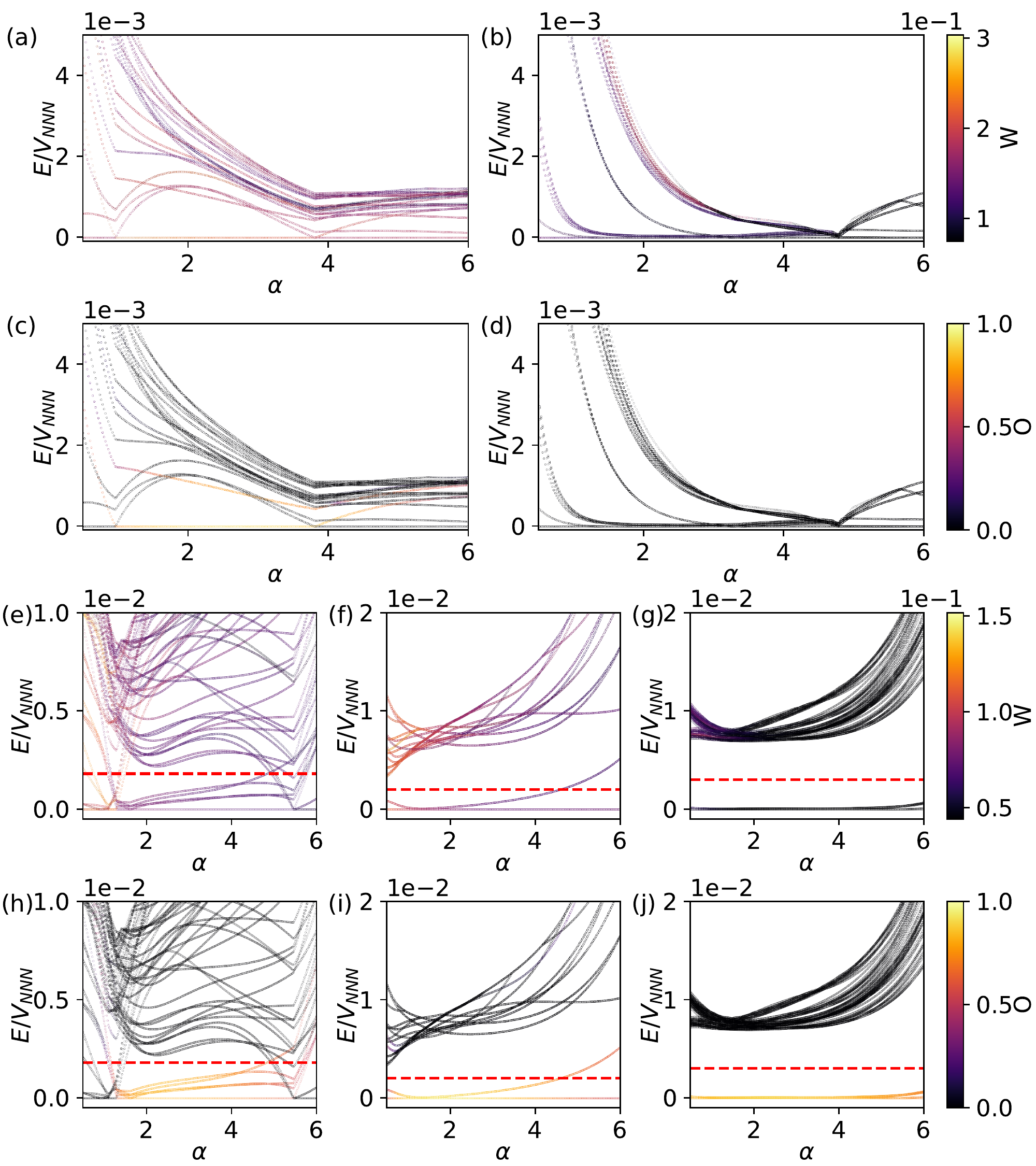}
    \caption{The energy spectra of $\nu=1/9$ $C=2$ systems as a function of $\alpha$. The colors in the rows 2,~4 and~1,~3 denote the Cartesian crystallization strength and the overlap with the model FQH wavefunctions, respectively. The first two rows correspond to the generalized Hofstadter model: (a),~(c) $ 4\times 9$, and (b),~(d) $6\times 6$. The next two rows contain the results for the triangular lattice model: (e), (h) $4\times 9$, and (f),~(i) $6\times 6$, and (g),~(j) $5\times 9$. We note that the energy levels are often degenerate, and these states can strongly differ in $W$ (i.e. each data point can coincide with an another one with lower or higher $W$).}
    \label{fig:C2_1_9}
\end{figure}

Fig.~\ref{fig:C2_1_9} shows the energy spectra color-coded with $W$ and overlap $O$ for these systems. Let us start by analyzing the overlaps. For all systems but one (Hofstadter~$6\times 6$), we observe the presence of nine states with quite high overlap with model FQH states (we have~$O>0.75$ in all these systems at some values of~$\alpha$). For each of these systems, there is a range of~$\alpha$ where these states are the lowest. Obviously, these states have also the same momenta as the model FQH ground states. Therefore, it seems that at these values of~$\alpha$ the systems is in the FCI phase.

\begin{figure}
    \centering
    \includegraphics[width=0.5\textwidth]{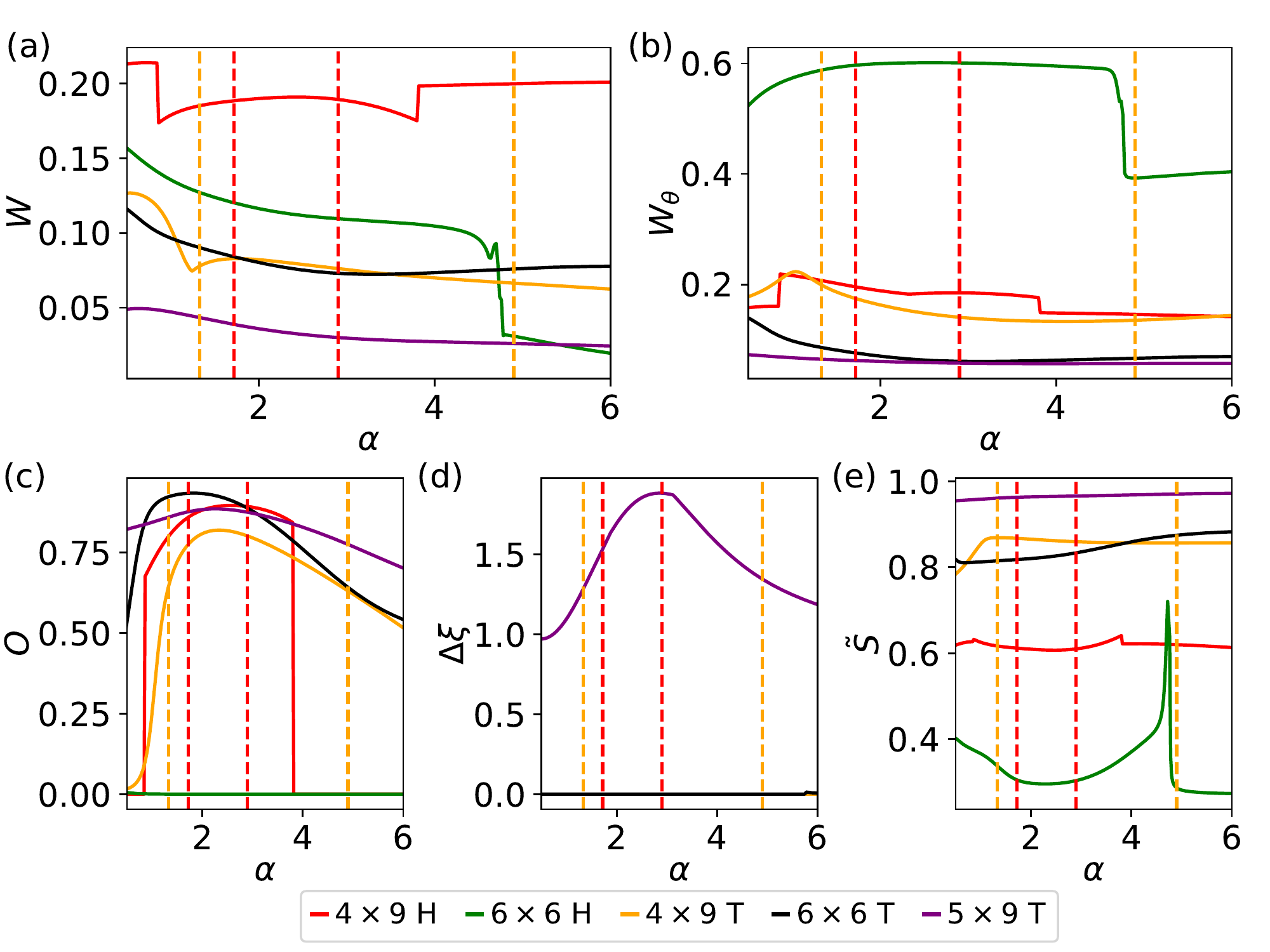}
    \caption{The FCI and WC characteristics as a function of $\alpha$ for triangular lattice~(T) and Hofstadter model~(H) and plaquettes at filling~$\nu=1/9$. The different plaquettes sizes are denoted by different colors.
     The upper row corresponds to the Wigner crystallization strength, polar~(a) and Cartesian~(b).
     The lower row corresponds to the FCI indicators: overlap with the FQHE state~(c), gap in the entanglement spectrum~(d), renormalized entanglement entropy $\tilde{S}=(S-S_{\mathrm{min}})/(S_{\mathrm{max}}-S_{\mathrm{min}})$~(e). The results are plotted for a state chosen using the procedure described in the main text. The dashed vertical lines correspond to the locations of gap closings.}
    \label{fig:WC_FCI_19_combine}
\end{figure}

Further WC and FCI characteristics for selected states are shown in Fig.~\ref{fig:WC_FCI_19_combine}. The procedure of choosing the state is similar as in Fig.~\ref{fig:WC_FCI_17_all} and~\ref{fig:WC_FCI_15_combine}, but we have to adjust it for two reasons. First, the absolute ground state at low~$\alpha$ does not always lie in a subspace consistent with FCI counting rules. Therefore, we focus only on the momenta corresponding to model FQH ground states. From these subspaces, we choose the ones where, at~$\alpha=0.5$, the lowest state has the lowest energy. Typically, this energy level is exactly degenerate, as degeneracy is common in the~$\nu=1/9$ case. Therefore, there are several such subspaces. Secondly, unlike the~$\nu=1/5$ cases, the degenerate states can differ significantly in crystallization strengths (i.e. the data points in rows~1 and~3 of Fig.~\ref{fig:C2_1_9} can coincide with ones with higher or lower $W$).  Therefore, among the selected subspaces, we choose the one in which the lowest state at low~$\alpha$ has highest~$W$. Then, we plot all the characteristics for the lowest state of this subspace for all~$\alpha$. 

From Fig.~\ref{fig:WC_FCI_19_combine}~(d) one can see that only the~$5\times 9$ triangular lattice system displays an entanglement gap. Similarly, the entanglement entropy of the selected state is closest to $S_{\mathrm{max}}$ for this system (Fig.~\ref{fig:WC_FCI_19_combine}~(e)). In contrast, for the $4\times 9$ Hofstadter plaquette the entanglement entropy is far from the maximal value ($\tilde{S}=(S-S_{\mathrm{min}})/(S_{\mathrm{max}}-S_{\mathrm{min}})\approx 0.6$. This suggests that the FCI state is weaker and less stable than in the cases studied previously.

Moreover, the behaviour of the FCI phase differs qualitatively between systems, as can be seen in Fig.~\ref{fig:C2_1_9}. In the~$6\times 6$ and~$5\times 9$ triangular lattice systems (Fig.~\ref{fig:C2_1_9}~(i),~(j)) the energy gap above the FCI ground-state manifold remains open in the entire investigated range of $\alpha$. In the~$4\times 9$ systems of both lattices (Fig.~\ref{fig:C2_1_9}~(c),~(h)) we observe two~$\alpha$ values where the gap closes -- an upper and lower limit to the FCI phase (note that the upper limit was not observed for the systems investigated previously in this work). Moreover, in the~$4\times 9$ Hofstadter system the gap is very small compared to the energy splitting of the ground-state manifold. In the~$6\times 6$ Hofstadter system (Fig.~\ref{fig:C2_1_9}~(d)), there is no FCI phase. Qualitative differences between systems can also be seen in Fig.~\ref{fig:WC_FCI_19_combine}, where the curves of FCI and WC characteristics can have significantly different shape for different plaquettes. These differences might be another signature of the fact that the FCI phase is weak and unstable, but may also have geometrical reasons -- the shape of our systems varies strongly. We study the plaquettes with aspect ratio~1 or close to~1 ($6\times 6$~triangular, $4\times 9$~Hofstadter -- remember that the Hofstadter unit cell has aspect ratio~3), as well as elongated ones ($6\times 6$~Hofstadter, $4\times 9$~and $5\times 9$~triangular). 

In addition to the FCI, we also observe Wigner crystals. By investigating $W$ and~$W_\theta$ (rows~1 and~3 of Fig. ~\ref{fig:C2_1_9} and Fig.~\ref{fig:WC_FCI_19_combine}~(a),~(b)), as well as inspecting the PCDs visually, we find that all the systems exhibit some form of crystalline order at low~$\alpha$. This happens even for the~$5\times 9$ triangular system, where $W$ is low for all states, and all values of~$\alpha$. In this case the crystal is weak (i.e. the particles are not as localized as in Fig.~\ref{fig:wc}~(b)) and deformed (i.e. the PCD maxima are displaced from their ideal periodic position), which may be a reason for low values of~$W$. Nevertheless, there are four PCD maxima, signifying the localization of particles, and the formation of the crystal coincides with a slight increase in the $W$.

If we define the phase transition point as the FCI gap closing, then such transitions exist only in the $4\times 9$ systems, at $\alpha\approx 1.72$, $\alpha\approx 2.90$ (Hofstadter) and $\alpha\approx 1.33$, $\alpha\approx 4.90$ (triangular). Nevertheless, as noted above, we observe some form of crystalline order also for other triangular lattice systems. In the cases investigated before within this work, we observed that the crystalline order starts to develop already at~$\alpha$ higher than the gap closing. This may also be the case here, i.e. the gap might close at $\alpha<0.5$. Another possibility is that the gap remains open due to the finite-size effects, and will close in the thermodynamic limit (provided that neither WC nor FCI disappears in infinite systems).

Another interesting case is the $4\times 9$ Hofstadter system. Figs.~\ref{fig:C2_1_9}~(a) and~\ref{fig:WC_FCI_19_combine}~(a) show that $W$ remains high even when the system is in the FCI phase, while, as noted before, the entanglement entropy is far from~$S_{\mathrm{max}}$ (Fig.~\ref{fig:WC_FCI_19_combine}~(e)). The visual inspection of the PCD shows that it displays crystalline order, i.e. the system simultaneously exhibits characteristics of WC and FCI. This is in line with the suggestion by Yang et al. \cite{HaldaneED}, (discussed also in Sec.~\ref{ssec:1over7transition}), that the FQH states have some crystal-like correlations built in. Because of such effects, the gap closing is not necessarily a good definition of transition point for the $C=2$ $\nu=1/9$. 

The differences between the systems studied in this subsections make any extrapolation to the thermodynamic limit even less reliable than for other cases considered in this work. The importance of the geometric effects can be seen when one compares the $4\times 9$ and $6\times 6$ plaquettes of the same lattice model - even though the number of sites is the same in both cases, and both have $N_{\mathrm{part}}=4$, the difference in aspect ratio leads to significantly different behaviuour of the two systems.

In summary, both Wigner crystal and FCI can exist at~$\nu=1/9$ of $C=2$ bands (at least in finite-size systems), and the transition between them can be triggered by controlling the interaction range. However, there are strong, qualitative differences in the behaviour of these phases in systems of various size, shape and underlying lattice model. Moreover, by combining these findings with results from Sec.~\ref{ssec:1_5}, we conclude that forming a Wigner crystal is harder in $C=2$ bands than in $C=1$ ones, in the sense that one has to consider lower filling factors. In other words, the~$C=2$ FCIs seem to be more stable against WC formation than their $C=1$ counterparts. The difference between $C=2$ and~$C=1$ is striking, compared to the small difference between the~$C=1$ and~$C=0$ cases reported in \cite{TFBWigner}, although we note that the comparison in Ref.~\cite{TFBWigner} was made for~$\alpha$ too small for the FCI to be observed at~$C=1$.

\section{Conclusions}\label{sec:conclusions}
In this work, we performed a finite-size exact-diagonalization study of transition between the FCI and Wigner crystal as a function of interaction range in the~$C=1$ and~$C=2$ flat-band lattice models. 

First, we studied the example of the $C=1$ band of the kagome lattice at~$\nu=1/7$. We analyzed five different characteristics of FCI and WC, all leading to the same conclusion: the FCI and WC emerge respectively for short- and long-range interaction, and hence it is possible to trigger a WC-FCI transition by controlling the interaction range. The results were qualitatively similar for three investigated systems. 

Next, to see how the WC formation is affected by band topology, we compared the behaviour of~$C=1$ and~$C=2$ models at~$\nu=1/5$. The former displayed an FCI-WC transition, although the WC was weaker than for~$\nu=1/7$. In the latter, however, the WC was absent, which suggests that the $C=2$ FCIs are more stable against the crystal formation than their $C=1$ counterparts at the same filling.

Finally, we studied the $C=2$ models at~$\nu=1/9$. In such a case, we observed both FCI and WCs, suggesting that one may be able to observe the FCI-WC transition for~$C=2$ systems. However, the behaviour of these phases as a function of~$\alpha$ exhibited significant, qualitative differences between the lattice models and system sizes.

We note that for the systems whose size is small enough for exact diagonalization, the geometry of the system and the number of particles plays an important role, e.g. by limiting the possible Wigner crystals consistent with the periodic boundary conditions. This may be an explanation for qualitative differences between the systems. Therefore, our results, strictly speaking, can be applied to finite-size systems only, and while we can speculate about the thermodynamic limit, we cannot make any definite conclusion about it. However, working on few-particle systems, with similar number of particles as discussed in this work, might be a feasible way of creating an FCI in optical lattices \cite{hudomal2019bosonic}, and in such case one does not need to analyze the thermodynamic limit. While the periodic boundary conditions were chosen by us for computational convenience, we note that an optical-lattice realization of a fractional Chern insulator in a torus geometry was proposed \cite{kim2018optical}. However, typical schemes of creating an FCI in optical lattices consider short-range interaction\cite{sorensen2005fractional,palmer2006high,palmer2008optical,hafezi2007fractional,opticalflux1}, so creating a tunable long-range interaction remains an experimental challenge.

 Further exploration of the transition for larger systems (perhaps also with open boundary conditions) may be performed using the DMRG method \cite{Johannes} or using model wavefunctions \cite{wu2013bloch,tu2014lattice}, as these methods were successful in investgating the WC-to-FQH transition in Landau levels \cite{MakiZotos, LamGirvin,ShibataYoshiokaDMRG}.


\begin{acknowledgments}
We thank Pawel Potasz and Alina Wania Rodrigues for helpful comments and careful proofreading of the manuscript.  B. J. thanks Zhao Liu for fruitful discussions regarding the implementation of the overlap computations.
M. K. was supported by the National Science Centre (NCN, Poland) under grant: 2019/33/N/ST3/03137. 
B. J. was supported by Foundation for Polish Science (FNP) START fellowship no.  032.2019 and Independent Research Fund Denmark under Grant Number 8049-00074B.
Our calculations were performed at the Wrocław Center for Networking and Supercomputing.
\end{acknowledgments}


\appendix

\section{Choice of kagome lattice parameters}\label{app:params}

The stability of the FCI phase depends not only on the many-body interaction, but also on the lattice parameters. 
In most cases, we have used well-known parameters from the literature \cite{wang2012fractional,wang2013tunable,Zoology,Tang}. The only exception is the kagome lattice with filling $\nu=1/7$, for which the FCI phase was not stable enough for various considered system sizes and interaction parameters.

\begin{figure}
    \centering
    \includegraphics[width=0.5\textwidth]{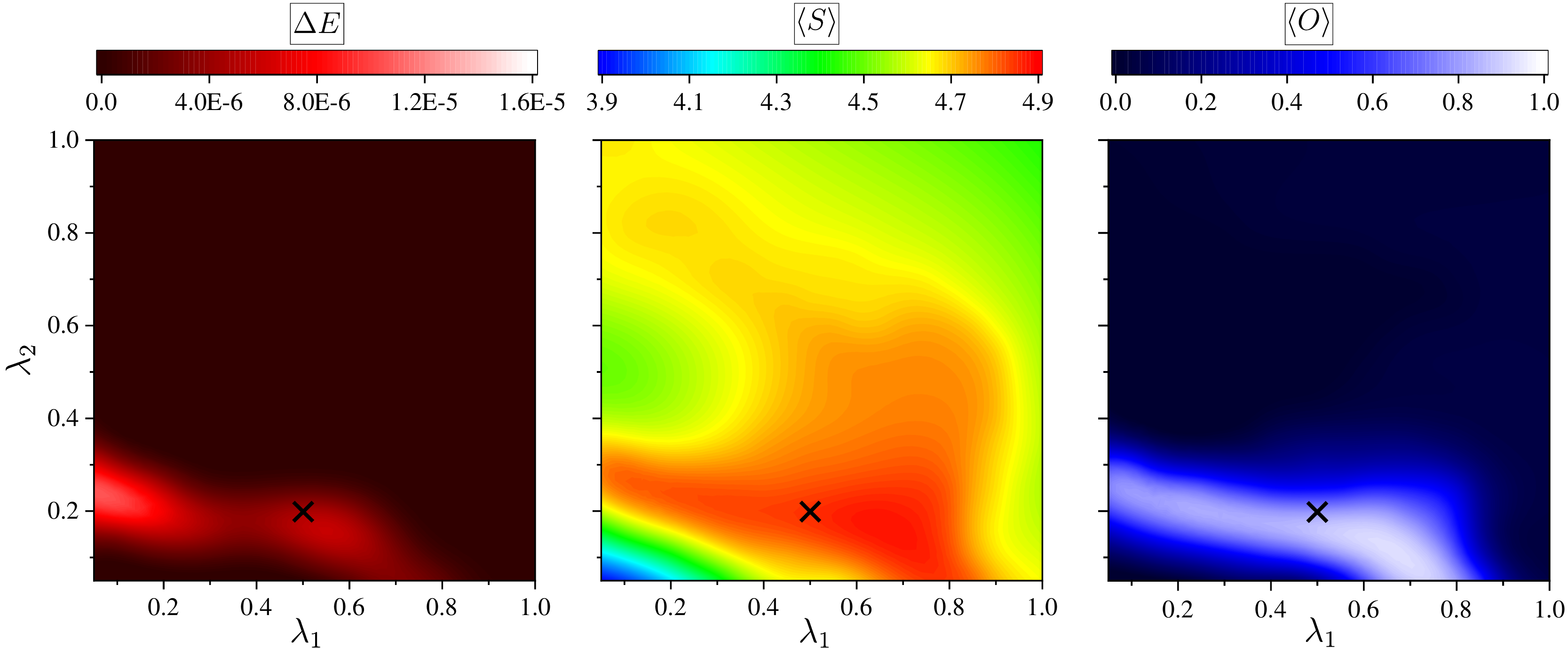}
    \caption{ Energy gap~$\Delta E$, average entanglement entropy~$\langle S \rangle$, average overlap of the FCI state with the FQHE state~$\langle O \rangle$,  respectively. Signatures of the FCI phase on the~$4\times 7$ plaquette with filling~$\nu=1/7$ as the function of the lattice parameters $\lambda_1$,~$\lambda_2$ with fixed values $t_1=1$,~$t_2=-0.3$, screened Coulomb interaction with~$\alpha=6.0$. }
    \label{fig:kagome_map_4x7}
\end{figure}

To determine more suitable values of parameters, we calculated signatures of the FCI phase as the function of the Hamiltonian~(\ref{eq:hkag}) parameters~$\lambda_1$ and~$\lambda_2$ with fixed values $t_1=1$,~$t_2= -0.3$. 
We focused on the~$4\times 7$ plaquette with $N=4$ particles with the screened Coulomb interaction in short range limit (screening parameter $\alpha = 6.0 $), see Fig.~\ref{fig:kagome_map_4x7}. 
The simplest signature of the FCI state is the energy gap~$\Delta E$ between the 7-fold quasi-degenerate FCI ground state at lattice momenta~$\mathbf{K}\in\{(2,0),(2,1),(2,2),(2,3),(2,4),(2,5),(2,6)\}$ and the first excited state. 
If one of the six lowest states has a momentum which does not belong to this set, i.e. the generalized Pauli principle is not fulfilled, we set~$\Delta E = 0$. 
Additionally, we calculated average particle entanglement entropy~$\langle S \rangle$ and average overlap with FQHE state~$\langle O \rangle$. 
The average is taken over seven states with the lowest energy.
By looking for that three signatures of the FCI state, we chose $\lambda_1=0.5$ and~$\lambda_2=0.2$.
These parameters are marked by the $ \times $ sign on the Fig.~\ref{fig:kagome_map_4x7}. We note here, that the question what is the influence of single-particle parameter on the stability of WC and properties of WC-FCI transition is still open.

\section{Signatures of the Wigner crystal -- details}\label{app:cryst_strength}
Here we provide a more detailed summary of the definitions of crystalization strength. For even more details, see \cite{TFBWigner}.

The PCD is turned into a continuous quantity by replacing every site with a Gaussian,
\begin{equation}
G_{i}(\mathbf{r})=\sum_{j=1}^{N} G(i,j)\frac{1}{\sigma\sqrt{2\pi}}\exp\left(-\frac{|\mathbf{r}-\mathbf{r_{j}}|^2}{2\sigma}\right),
\label{eq:pcdcont}
\end{equation}
where $\sigma$ is the width of the Gaussian, and~$\mathbf{r}$ is the vector connecting site~$i$ (where the fixed particle is located) and a given point in space. We use~$\sigma=0.5$. Typically, the results do not differ significantly for different starting sites~$i$, hence we can choose any site and drop this index, provided that we measure $\mathbf{r}$ with respect to that site. 

To obtain the Cartesian Fourier transform, we first discretize this continuum PCD on a regular grid~$N_{\mathrm{grid,1}} \times N_{\mathrm{grid,2}}$, obtaining a matrix
\begin{equation}
    \tilde{G}_{mn}=G\left(\frac{mN_1}{N_{\mathrm{grid,1}}}\mathbf{a}_1+\frac{nN_2}{N_{\mathrm{grid,2}}}\mathbf{a}_2\right),
\end{equation}
where $\mathbf{a}_1$, $\mathbf{a}_2$ are the lattice vectors of the tight-binding model. We perform a discrete Fourier transform of~$\tilde{G}_{mn}$ using the FFT algorithm and obtain the Fourier coefficients $F_{mn}$, which we normalize by dividing by the magnitude of the zeroth component $\tilde{F}_{mn}=F_{mn}/|F_{00}|$. Thus, the pair correlation density in the momentum space is given, up to normalization, by
\begin{equation}
    F_{c}(\mathbf{k})=\sum_{mn} \tilde{F}_{mn}\delta(\mathbf{k}-\frac{m}{N_1}\mathbf{b}_1-\frac{n}{N_2}\mathbf{b}_2),
    \label{eq:pcd_reciprocal}
\end{equation}
where $\mathrm{c}$ denotes ``Cartesian'', and  $\mathbf{b}_1$,~$\mathbf{b}_2$ are the reciprocal lattice vectors obtained from~$\mathbf{a}_1$,~$\mathbf{a}_2$. The Fourier peaks located at~$m, n$ being multiples of~$N_1,N_2$, respectively corresponds to the reciprocal lattice of the tight-binding model. Any~$m, n$ smaller than~$N_1,N_2$, respectively, are responsible for features varying on a scale larger than a single unit cell, i.e. a possible Wigner crystal. However, not every such pattern is a WC: it needs to be periodic in two directions, i.e. two Fourier components should be nonzero, and the number of maxima of the corresponding real-space pattern should match the number of particles. Thus, for every~$N_{\mathrm{part}}$ there is a finite number~$N_{W}$ of possible Wigner crystals, each labelled with two integer vectors~$(m_i, n_i), (o_i, p_i)$. We determine them by listing all possible combinations of these integers, and neglecting all these with incorrect number of maxima. Then, for every possible crystal, we calculate the corresponding crystalization strength by multiplying the two Fourier components described by these vectors. Next, we take the maximum value of this product over all crystals as the crystalization strength (Eq.~\eqref{eq:cryst_strength}). We note that $\tilde{G}_{mn}$~does not have the exact periodicity of the reciprocal lattice of the Wigner crystal, as the ``hole'' at the position of the fixed particle breaks the periodicity of the original PCD (see the Appendix A.3 of~\cite{TFBWigner} for details). This can generate nonzero~$W$ for a non-crystalline PCD pattern, but, compared to~$W$ for a Wigner crystal in the same system, it is generally smaller.

We can also perform the angular Fourier transform, which we do by discretizing~$G(\mathbf{r})$ on a polar grid and performing FFT at each $r$ separately. This yields the $r$-dependent Fourier coefficients~$F_{\theta} (r, k_\theta)$. As noted in the main text, we look at~$k_\theta=2,4,6$, related to 2,4,6-fold rotational symmetry. However, we should bear in mind that PCD is periodic with plaquette periodicity. That is, even if PCD is uniform far away from the fixed particle, the periodic images of the ``hole'' around its position will introduce an artificial angular periodicity. Thus, we need to introduce a cutoff radius~$r_{\mathrm{max}}$. As a compromise between avoiding the ``holes'' and capturing as many particles as possible, we choose~$r_{\mathrm{max}}$ equal~0.6 times the distance to nearest periodic image of the fixed particle. Having this in mind, we define the normalized angular Fourier transform as
\begin{equation}
    \tilde{F}_{\theta} (r, k_\theta)=\frac{F_{\theta} (r, k_\theta)}{\max_{r<r_{\mathrm{max}}} |F_{\theta}(r,0)|}.
\end{equation}
Then we proceed as described in Sec. \ref{ssec:WCsignatures}.

\section{Entanglement signatures of FCI}\label{app:entropy}
The existence of the FCI phase can be seen using entanglement methods. Here, we focus on the particle partition. Typically in the FCI literature \cite{sterdyniak2011extracting, PRX}, one constructs a density matrix as an equal-weight superposition of the pure-state density matrices of all $q$ quasi-degenerate ground states,
\begin{equation}
    \rho=\frac{1}{q}\sum_{i=1}^{q}\ket{\psi_i}\bra{\psi_i}.
    \label{eq:rho}
\end{equation}
Then, one divides the system into two subsystems $A$ and $B$, with $N_A$ and $N_B$ particles ($N_A+N_B=N_{\mathrm{part}}$), and performs a trace over the $B$ subsystem, $\rho_A=\mathrm{Tr}_B\rho$. From the eigenvalues $\lambda_i$ of the reduced density matrix $\rho_A$ one constructs the entanglement energies $\zeta_i=-\ln \lambda_i$. 

In our work, we follow this approach, but instead of using $\rho$ defined by Eq.~\eqref{eq:rho} we construct a pure-state density matrix of each energy eigenstate separately
\begin{equation}
    \rho_i=\ket{\psi_i}\bra{\psi_i}.
    \label{eq:rhoi}
\end{equation}
This definition also works for~$i>q$, i.e. the excited states.

As we noted in the main text, in most of the studied cases even such a single-state entanglement spectrum displays the gap and correct counting of states below it for FCI. We define the entanglement gap in the following way. Let~$n_{\mathrm{P}}(K_1, K_2)$ be the number of entanglement energy levels consistent with the generalized Pauli principle \cite{PRX,bernevig2012emergent} in the~$\mathbf{K}=[K_1,K_2]$ subspace. We denote the $i$th entanglement energy level in the~$\mathbf{K}$ subspace as~$\zeta_{i}(K_1,K_2)$, (we assume that they are sorted in an increasing order, i.e. $\zeta_{i}(K_1,K_2)\leq \zeta_{j}(K_1,K_1)$ for~$i<j$). We define two sets of entanglement energies, $\zeta_{\mathrm{below}}=\{\zeta_{i}(K_1,K_2): i\leq n_{\mathrm{P}}(K_1,K_2), K_1=0,\dots N_1-1, K_2=0,\dots N_2-1\}$ and
$\zeta_{\mathrm{above}}=\{\zeta_{i}(K_1,K_2): i>n_{\mathrm{P}}(K_1,K_2), K_1=0,\dots N_1-1, K_2=0,\dots N_2-1\}$. The entanglement gap is defined as
\begin{equation}
    \Delta \zeta =\max\left\{0, \min\left\{\zeta_{\mathrm{above}} \right\}-\max\left\{\zeta_{\mathrm{below}} \right\} \right\}. 
\end{equation}
In the case of~$C=2$ states, instead of implementing the generalized Pauli principle, we compare the entanglement spectra of the topological flat band systems, obtained using \eqref{eq:rhoi}, to the entanglement spectra of the model continuum Halperin-like states, computed using \eqref{eq:rho}. That is, $n_{\mathrm{P}}(K_1, K_2)$ used in the calculation of $\Delta \zeta$ for the investigated systems is the number of entanglement energy levels below the gap in the corresponding model state. 

Instead of looking at the structure of the entanglement spectrum, we can use the entanglement entropy,
\begin{equation}
    S=- \sum_i\lambda_i \ln \lambda_i.
\end{equation}
In Ref. \cite{haque2007entanglement, Zozulya2}, an exact upper bound for the entanglement entropy of Laughlin states was obtained,
\begin{equation}
    S_{\mathrm{max}}=\ln \sum_{K_1,K_2}n_{\mathrm{P}}(K_1, K_2).
\end{equation}
It is derived by assuming that all $\zeta_i$ below the gap are equal, and all the others are infinite, i.e. the corresponding~$\lambda_i$ equal~0 and do not contribute to the entropy. The authors of Refs. \cite{haque2007entanglement, Zozulya2} found numerically that the entanglement entropy of the continuum Laughlin states is close to that bound. We expect that the same will happen for FCI, both at~$C=1$ and~$C=2$. 

There is also a lower bound on the entanglement entropy, obtained by assuming that the state~$\ket{\psi_i}$ is a single Slater determinant,
\begin{equation}
    S_{\mathrm{min}}=\ln  {{N_{\mathrm{part}}}\choose{N_A}}.
\end{equation}
We expect that the entanglement entropy of the FCIs will be far larger than this minimum value.

\section{Overlap with model FQH states}\label{app:overlap}

To calculate the overlap with a model wavefunction, three problems need to be solved. First of all, the FQH and FCI states should carry the same quantum numbers. That is, in the single-particle bases~$\ket{\phi_{\mathrm{FCI}}(\mathbf{k})}$,~$\ket{\phi_{\mathrm{FQH}}(\mathbf{k})}$ for FQH and FCI, the same sets of values of $\mathbf{k}=[k_1,k_2]$ momenta should be allowed.  If we fully exploit the translational symmetry of the FCI fully, we have~$k_1=0,\dots, N_1-1$ and~$ k_2=0,\dots, N_2-1$ as allowed momenta. However, in the Landau gauge for the FQH systems, the Brillouin zone is different -- it is 1-dimensional, with~$k=0,1,\dots N_1N_2$, Therefore, another basis for FQH systems should be used.

The second problem is that on a torus, we can add a phase~$e^{i\gamma_1}$,~$e^{i\gamma_2}$ at the boundary conditions in the directions $\mathbf{a}_1$,~$\mathbf{a}_2$. After such a modification, the system stays in the FCI/FQH phase -- indeed, these phases are varied during the calculation of FCI/FQH signatures, such as spectral flow or many-body Chern number (see e.g. \cite{sflow1, Neupert}). These phases control the Berry phase of a particle encircling the torus around its fundamental cycles (large Wilson loops), and to maximize the overlap, we should demand that the respective large Wilson loops are equal for FCI and FQH. This does not necessarily mean that the boundary condition phases are equal for FQH and FCI. Therefore, we fix~$\gamma_1=\gamma_2=0$ for FCI and search for the optimal $\gamma_1$,~$\gamma_2$ in FQH.

Thirdly, we should specify the mapping between FQH and FCI precisely. To compute the overlap~$\braket{\psi|\psi_{FQH}}$ between the ED result and the model wavefunction, we need to know $\braket{\phi_{\mathrm{FCI}}(\mathbf{k})|\phi_{\mathrm{FQH}}(\mathbf{k}')}$, i.e. the overlap between the single-particle basis functions for FCI and FQH systems. Since  $\ket{\phi_{FCI}(\mathbf{k})}$,~$\ket{\phi_{\mathrm{FQH}}(\mathbf{k})}$ describe different systems, it is up to us to define the relation between them by fixing the values of~$\braket{\phi_{\mathrm{FCI}}(\mathbf{k})|\phi_{\mathrm{FQH}}(\mathbf{k}')}$. It is natural to identify the states with the same momenta, i.e. to set $\braket{\phi_{\mathrm{FCI}}(\mathbf{k})|\phi_{\mathrm{FQH}}(\mathbf{k}')}=0$ if $\mathbf{k}\neq \mathbf{k}'$. However, this still leaves us with some ambiguity. Let us assume that the basis~$\ket{\phi_{\mathrm{FCI}}(\mathbf{k})}$ are the lowest-band eigenfunctions resulting from the numerical diagonalization of the single-particle model. We can define a different basis for the FCI, by multiplying every basis vector by a momentum-dependent phase~$\ket{\tilde{\phi}_{\mathrm{FCI}}(\mathbf{k})}=e^{i\theta_{\mathbf{k}}}\ket{\phi_{\mathrm{FCI}}(\mathbf{k})}$. We can require either that $\braket{\phi_{\mathrm{FCI}}(\mathbf{k})|\phi_{\mathrm{FQH}}(\mathbf{k})}=1$ or $\braket{\tilde{\phi}_{\mathrm{FCI}}(\mathbf{k})|\phi_{\mathrm{FQH}}(\mathbf{k})}=1$. These two options result in two different values of the overlap. Therefore, we have to find a mapping for which~$\braket{\psi|\psi_{FQH}}$ is maximal. That is, given~$\ket{\phi_{\mathrm{FCI}}(\mathbf{k})}$, we have to find the phases $\theta_{\mathbf{k}}$ which transforms it into an another basis~$\ket{\tilde{\phi}_{\mathrm{FCI}}(\mathbf{k})}$, which maximizes~$\braket{\psi|\psi_{FQH}}$ under the condition~$\braket{\tilde{\phi}_{\mathrm{FCI}}(\mathbf{k})|\phi_{\mathrm{FQH}}(\mathbf{k})}=1$. This is what we mean by ``fixing the gauge''.

The solutions for all the three problems were given in Ref.~\cite{wu2013bloch}. The authors proposed a Bloch basis for FQH systems, indexed by a momentum in~$N_1\times N_2$ Brillouin zone, and an algorithm which provides appropriate $\theta_{\mathbf{k}}$, $\gamma_1$ and $\gamma_2$. The algorithm computes the Berry connection and the large Wilson loops in the FCI case and compares with the result for FQH, adjusting the~$\gamma_1$,~$\gamma_2$ accordingly. The gauge $\theta_{\mathbf{k}}$ is found by imposing a discrete analog of Coulomb gauge condition for the FCI and solving a discretized Poisson equation with Berry curvature fluctuation as a source. The algorithm was implemented in the DiagHam software \cite{diagham}, and in our work we use a DiagHam-based code to optimize the overlap. 

The basis~$\ket{\tilde{\phi}_{\mathrm{FCI}}(\mathbf{k})}$ is then fed to the ED calculation, i.e. the Hamiltonian~\eqref{eq:projint} is diagonalized in the many-particle basis constructed as Slater determinants of these wavefunctions. The model FQH states for the overlap are constructed by diagonalizing the appropriate pseudopotential Hamiltonian in the Bloch basis, taking into account the boundary condition phases $\gamma_1$,$\gamma_2$. This is true for both $C=1$ and $C=2$. In the latter case, a bilayer system is considered, with boundary conditions mixing the layers. The Bloch basis is constructed following Refs.~\cite{wu2013bloch, wu2014haldane}. The model wavefunctions for $\nu=1/5$ are obtained with two first pseudopotentials $V_0=V_1=1$ and the rest equal to zero, while for $\nu = 1/9$ four first pseudopotentials are equal to unity, and the rest is zero.

\section{System size analysis}\label{app:different_sizes}

In this appendix, we present the signatures of WC and FCI phases in the low-energy spectrum of the systems partially described in the main article.

\subsection{ $ C=1 $, $\nu=1/7$}
\label{app:1over7}
\begin{figure}
    \centering
    \includegraphics[width=0.5\textwidth]{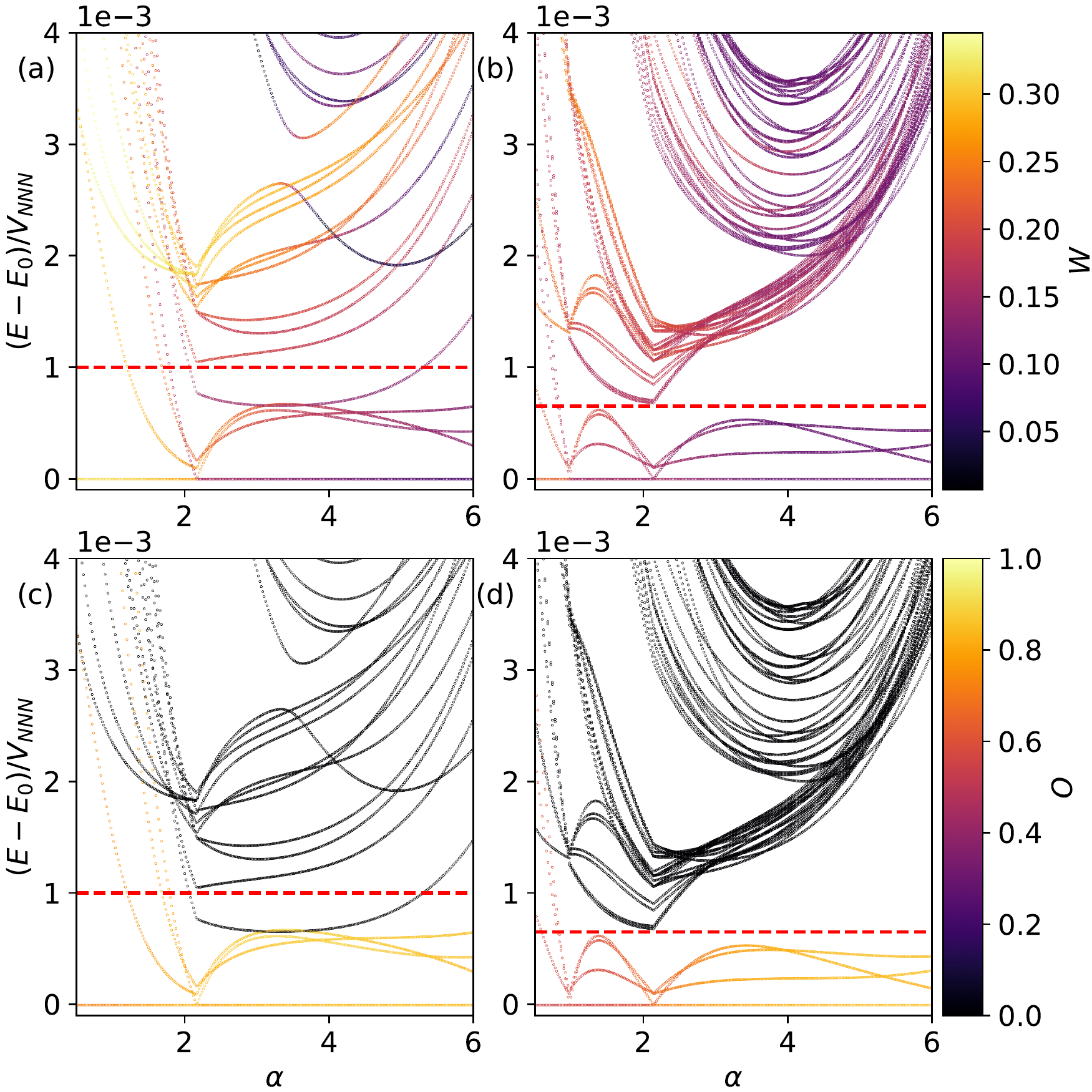}
    \caption{ The Cartesian strength of Wigner crystallization~$W$ (upper row), and overlap~$O$ with the model FQHE states (lower row) on the energy spectrum of the~$4\times 7$ (left column) and~$6\times 7$ (right column) kagome lattice at filling~$\nu =1/7$. 
    The overlaps in the momentum subspaces not corresponding to a model FQHE ground state are set to~0 by definition.  
    }
    \label{fig:different_sizes_1_7}
\end{figure}

 In section \ref{sec:1over7} we have shown the WC and FCI signatures for the kagome lattice system at filling~$\nu=1/7$, for the whole energy spectrum for the~$5\times 7$ plaquette and a single selected state for~$4\times 7$ and $6\times 7$ plaquettes.
The energy spectrum with the WC and FCI signatures for the last two systems is plotted in Fig.~\ref{fig:different_sizes_1_7}.
The Cartesian WC strength~$W$ is shown in Fig.~\ref{fig:different_sizes_1_7}~(a) and~(b), and the overlap with the model FQHE states is shown in Fig.~\ref{fig:different_sizes_1_7}~(c) and~(d).

The behaviour of these systems is similar to the~$5\times 7$ case described in Sec.~\ref{sec:1over7}, in the sense that for large~$\alpha$ we obtain an FCI with seven quasi-degenerate ground states, and as we lower $\alpha$ this ground-state manifold splits while the Wigner crystals emerge. 
The spectra from Fig.~\ref{fig:different_sizes_1_7} display also some differences with respect to the~$5\times 7$ case. 
For the~$4\times 7$ system, the gap above the FCI ground state manifold closes temporarily between~$\alpha\approx 3.1$ and~$\alpha\approx 3.6$.
Nevertheless, in that region, these states still have a large overlap with model FQHE states ($O>0.83$).
As for the~$6\times 7$ system, we can see that the gap closing occurs for much smaller~$\alpha$ than for two other systems. 
We also note that similarly to the~$5\times 7$ case, the Wigner crystals arise in the excited states as well, sometimes with larger~$W$ than in the ground state. Interestingly, in both the~$4\times 7$ and the~$6\times 7$ system, in some of the excited states we obtain a different Wigner lattice than in the ground states.

\subsection{$ C=1 $, $ \nu=1/5$}\label{app:1over5}
\begin{figure}
    \centering
    \includegraphics[width=0.5\textwidth]{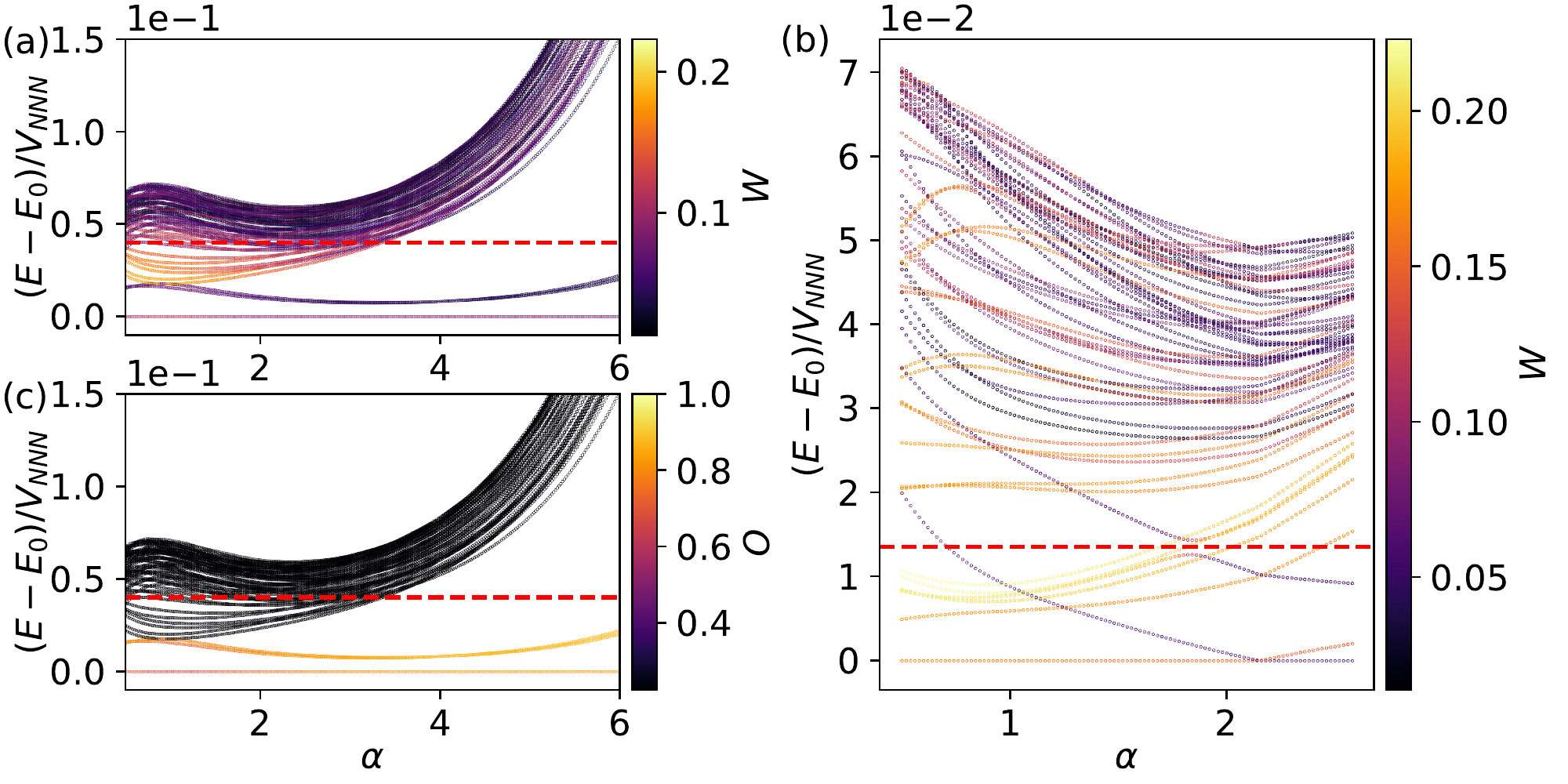}
    \caption{The Cartesian strength of Wigner crystallization~$W$ ((a)~and~(b)), and overlap~$O$ with the model FQHE state~(c) on the 
    the energy spectrum of the~$7\times 5$ ((a)~and~(c)) and~$8\times 5$ (b)~kagome lattice at filling $\nu =1/5$. 
    The overlaps in the momentum subspaces not corresponding to a model FQHE ground state are set to~0 by definition. }
    \label{fig:different_sizes_1_5}
\end{figure}

 In section \ref{ssec:1_5} we have shown the phase transition between WC and FCI phases on the kagome lattice system at filling~$\nu=1/5$, for the whole energy spectrum for the~$6\times 5$ plaquette and the one selected state for~$7\times 5$ and~$8\times 5$ plaquettes.
Fig. \ref{fig:different_sizes_1_5} (a) and (b) show the energy spectra as a function of~$\alpha$ color-coded with Cartesian crystallization strength for the plaquette~$7\times 5$ and~$8\times 5$, respectively.
The overlap with the FQHE state for the~$7\times 5$ system is shown in Fig.~\ref{fig:different_sizes_1_5}~(c).
 The results for the plaquette $8\times 5$ are obtained only in the limited range of $\alpha$, because of the numerical complexity of the computation and problems with numerical convergence of the diagonalization problem, especially in the limit of high~$\alpha$ values. For all systems at large~$\alpha$, we observe five quasi-degenerate ground states, which momenta match FCI counting rules.
Also, all ground states have crystalline order in the long-range-interaction limit.

From Fig. \ref{fig:different_sizes_1_5} and the consideration in section~\ref{ssec:1_5}, it can be seen that as we decrease $\alpha$, the gap above the FCI quasi-degenerate ground state manifold decreases and eventually closes. This process looks slightly different than in the case of~$\nu=1/7$: not all ground states cross before the gap closing occurs (e.g. in Fig.~\ref{fig:different_sizes_1_5}~(a) the absolute ground state does not cross with any other state all through the transition). As~$\alpha$ is lowered, the crystallization strength increases in one or two states which eventually become the absolute ground states. It is however interesting to note that in the~$7\times 5$ system, the crystallization strength in the ground state has a maximum at~$\alpha=0.91$. Upon further decrease of~$\alpha$, the crystallization strength drops, and at~$\alpha=0.5$ the Wigner crystal is nonexistent. This is an explicit example that a too-small screening can be detrimental for Wigner crystals. 
Similarly to the~$\nu = 1/7$ case, in a few excited state, the crystalline order exists and is visible even when the FCI phase is well established.  

\subsection{$C=2$, $\nu=1/5$}\label{app:1over5_2}
\begin{figure}
    \centering
    \includegraphics[width=0.5\textwidth]{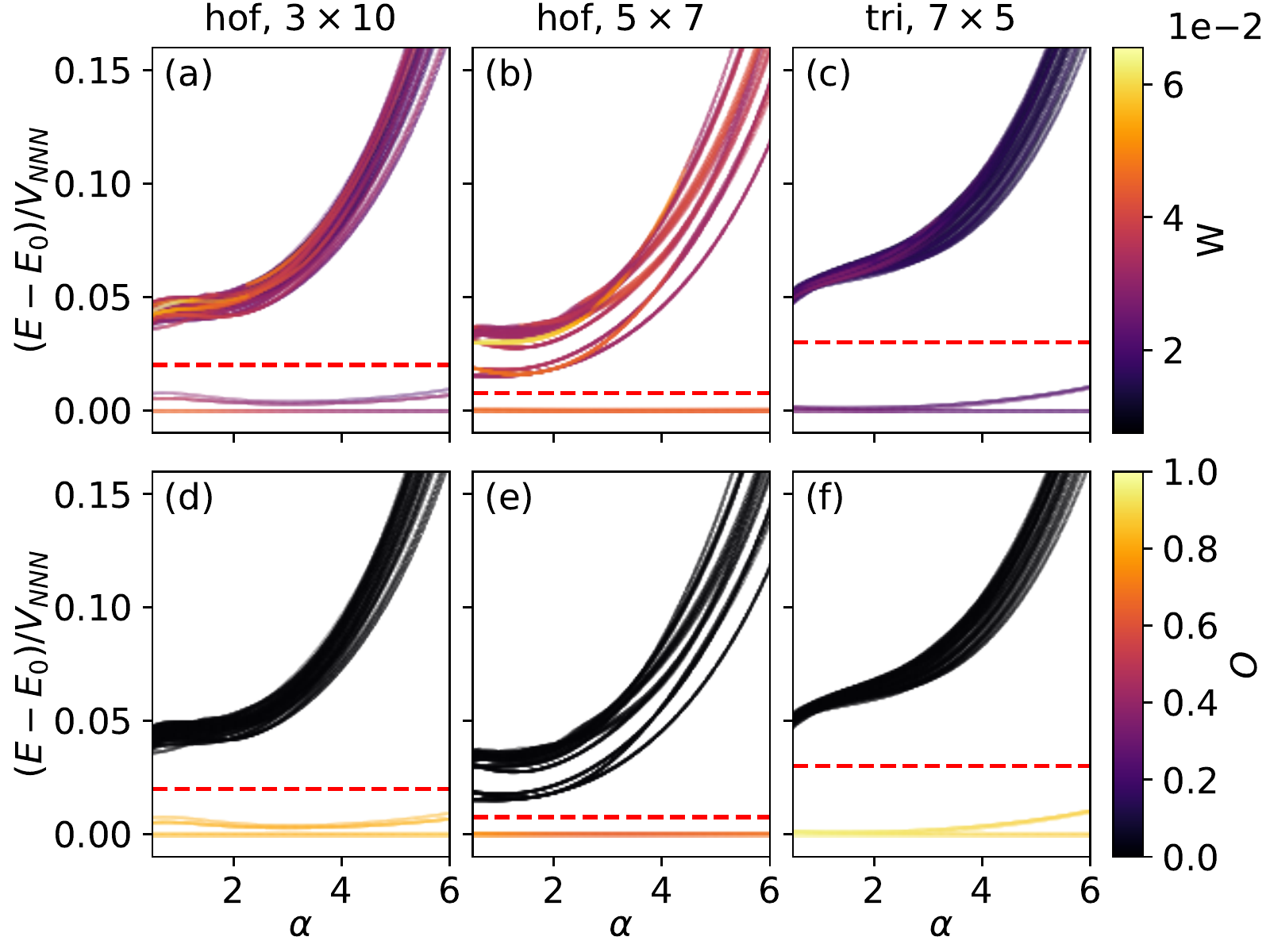}
    \caption{Signatures of WC and FCI on the low-lying energy spectrum for models with Chern number~$C=2$ with the filing factor~$\nu=\frac{1}{5}$ for different plaquettes: 
    Hofstadter model~(hof) $3\times 10$ (first column) and $5\times 7$ (second column), and triangular lattice~(tri) $7\times 5$ (third column).
   In the upper row, we plot the Cartesian crystallization strength~$W$, and the lower row displays the overlap~$O$ with model FQH states. The overlaps in the momentum subspaces not corresponding to a model FQHE ground state are set to~0 by definition.  }
    \label{fig:different_sizes_C2_1_5}
\end{figure}

In section~\ref{ssec:1_5} we have shown the phase transition between WC and FCI phases at filling~$\nu=1/5$ on the lattice systems with Chern number~$C=2$. 
The signatures of both phases have been plotted on the low-energy spectrum for the $6 \times 5$ Hofstadter lattice in the Fig.~\ref{fig:WC_FCI_15_compare} and for the one selected state in the Fig.~\ref{fig:WC_FCI_15_combine} for the following plaquettes: $3\times 10$, $5\times 7$ Hofstadter model and $7\times 5$ triangular lattice.
In the Fig.~\ref{fig:different_sizes_C2_1_5} is shown the full spectrum of the last mentioned plaquette with the color-coded Cartesian crystallization strength~$W$ (in Figs.~\ref{fig:different_sizes_C2_1_5}~(a),~(b),~(c)) and the overlap with the FQH states ( in Figs.~\ref{fig:different_sizes_C2_1_5}~(d),~(e),~(f)).
The behaviour of each system is similar: in the entire studied range of~$\alpha$ a five-fold quasi-degenerate ground state manifold is separated by a gap from the rest of the spectrum (see the dashed red horizontal line in Fig.~\ref{fig:different_sizes_C2_1_5}).
The momenta of these states are the same as the momenta of the model FQH system, and the overlap with them remains large (the lower row of Fig.~\ref{fig:different_sizes_C2_1_5}).
Also, the crystallization strength of the ground state (the upper row of Fig.~\ref{fig:different_sizes_C2_1_5}) remains small for all~$\alpha$ values for all states, but a few excited states are characterized by a larger value of the WC strength~$W$ in the small values of~$\alpha$. 

\bibliography{doktorat,fci,fqhe,bib}

\end{document}